\documentclass[11pt,a4paper]{article}
 \pdfoutput=1

\usepackage{amsmath,amssymb}
\usepackage{caption}
\usepackage{graphicx}
\usepackage{dcolumn}
\usepackage{bm,xcolor}
\usepackage{hyperref}
\usepackage{accents}
\usepackage{amssymb,float}
\usepackage{amsmath}
\usepackage{multirow}
\usepackage{siunitx}
\usepackage{tabularx}
\usepackage{booktabs}
\usepackage{authblk}
\usepackage{url}
\usepackage{geometry}

\usepackage{transparent}
\usepackage{tikz}

\DeclareSIUnit\parsec{pc} 
\hypersetup{
    colorlinks=true,
    linkcolor=red,
    filecolor=magenta,      
    citecolor=blue
}

\newcommand{\udt}[3]{#1^{#2}_{\phantom{#2}#3}}

\newcommand{\dut}[3]{#1_{#2}^{\phantom{#2}#3}}

\newcommand{\lc}[1]{\accentset{\circ}{#1}}%Levi-Civita connection

\title{Constraints on \texorpdfstring{$f(T)$}{f(T)} Cosmology with Pantheon+}

\author[1,2]{Rebecca Briffa, \thanks{\href{rebecca.briffa.16@um.edu.mt}{rebecca.briffa.16@um.edu.mt}}}
\author[3]{Celia Escamilla-Rivera,\thanks{\href{celia.escamilla@nucleares.unam.mx}{celia.escamilla@nucleares.unam.mx}}}
\author[1,2]{Jackson Levi Said, \thanks{\href{jackson.said@um.edu.mt}{jackson.said@um.edu.mt}}}
\author[1,2]{and Jurgen Mifsud \thanks{\href{jurgen.mifsud@um.edu.mt}{jurgen.mifsud@um.edu.mt}}}

\affil[1]{Institute of Space Sciences and Astronomy, University of Malta, Malta, MSD 2080}
\affil[2]{Department of Physics, University of Malta, Malta}
\affil[3]{Instituto de Ciencias Nucleares, Universidad Nacional Aut\'{o}noma de M\'{e}xico, Circuito Exterior C.U., A.P. 70-543, M\'exico D.F. 04510, M\'{e}xico}

\definecolor{mygray}{gray}{0.6}
\definecolor{myorange}{HTML}{F97306}
\definecolor{patriarch}{HTML}{7E1E9C}
\definecolor{myred}{HTML}{E50000}
\title{Growth of structures using redshift space distortion in \texorpdfstring{$f(T)$}{f(T)} Cosmology}

\date{\today}

\begin{document}

\maketitle
\abstract{
%\cel{In the Title: Cosmology or Cosmologies? or gravity?}\jls{We updated the title but it is ok as collective singular, I think}
Cosmology faces a pressing challenge with the Hubble constant ($H_0$) tension, where the locally measured rate of the Universe's expansion does not align with predictions from the cosmic microwave background (CMB) calibrated with $\Lambda$CDM model. Simultaneously, there is a growing tension involving the weighted amplitude of matter fluctuations, known as $S_{8,0}$ tension. Resolving both tensions within one framework would boost confidence in any one particular model. In this work, we analyse constraints in $f(T)$ gravity, a framework that shows promise in shedding light on cosmic evolution. We thoroughly examine prominent $f(T)$ gravity models using a combination of data sources, including Pantheon+ (SN), cosmic chronometers (CC), baryonic acoustic oscillations (BAO) and redshift space distortion (RSD) data. We use these models to derive a spectrum of $H_0$ and $S_{8,0}$ values, aiming to gauge their ability to provide insights into, and potentially address, the challenges posed by the $H_0$ and $S_{8,0}$ tensions.
}

\section{\label{sec:intro}Introduction}

Measurements of the accelerating expansion of the Universe \cite{SupernovaSearchTeam:1998fmf,SupernovaCosmologyProject:1998vns} have led to the prospect that it may be expanding faster than predicted by the $\Lambda$CDM model \cite{DiValentino:2021izs}. This may open the possibility that the concordance model description of gravitation through general relativity (GR), the as yet unobserved cold dark matter (CDM) \cite{Baudis:2016qwx,Bertone:2004pz,Gaitskell:2004gd}, and the theoretically problematic cosmological constant \cite{Peebles:2002gy,Copeland:2006wr} may require additions or corrections to its explanation of some kind. Over the last few years this has been expressed primarily through the value of the Hubble constant \cite{DiValentino:2020zio} and $f\sigma_{8,0}$ \cite{DiValentino:2020vvd}. The appearance of cosmic tensions has shaped into a tension between observations based on direct measurements of the late Universe such as those based on Type Ia supernovae, the tip of the red giant branch measurements, strong lensing measurements \cite{Riess:2021jrx,Wong:2019kwg,Anderson:2023aga,Freedman:2020dne}, and indirect measurements coming from analyses of early time data from the cosmic microwave background radiation (CMB) as well as big bang nucleosynthesis (BBN) data \cite{Aghanim:2018eyx,DES:2021wwk,eBOSS:2020yzd,Zhang:2021yna,Cooke:2017cwo} and others \cite{Benisty:2023vbz}. Furthermore, recent data forecasting from new missions that include structure formation have increased this tension at local level \cite{Atek_2022,Lu:2022utg}, however, considering new systematic techniques on the measurements these tensions can be treated \cite{Alonso:2023oro}.
The growth in both these tensions has led to a reinvigorated exploration of possible modifications of the concordance model that have been developed in the literature over the last few decades.

The literature features a myriad of novel approaches in which to confront the growing issue of cosmic tensions. There have been proposals involving a reevaluation of the cosmological principle \cite{Krishnan:2021jmh,Krishnan:2021dyb}, possible impacts of early Universe dark energy \cite{Poulin:2023lkg}, the appearance of extra degrees of freedom in the form of additional neutrino species in the early Universe \cite{DiValentino:2021imh,DiValentino:2021rjj}, as well as modifications to gravity \cite{CANTATA:2021ktz,AlvesBatista:2021eeu,Barack:2018yly,Abdalla:2022yfr} and others \cite{Addazi:2021xuf}. Many of these approaches modify a small part of the evolution of the Universe using new physics. On the other hand, modified gravity has the potential to make smaller changes that infiltrate the larger range of redshifts. Moreover, modifications to GR will also provide changes both at background and perturbative levels. In the space of possible modifications to GR, one approach that has become popular in recent years and which natively builds a competitive way in which to consider new physics is metric-affine gravity which is based on the exchange of the underlying connection with other possible geometries \cite{BeltranJimenez:2019esp,Hehl:1994ue}. In teleparallel gravity (TG), the geometric curvature of the Levi-Civita connection $\lc{\Gamma}^{\sigma}{}_{\mu\nu}$ (over-circles denotes quantities based on the curvature of the Levi-Civita connection in this work) is interchanged with the torsion of the teleparallel connection $\Gamma^{\sigma}{}_{\mu\nu}$. This may provide a more intuitive approach in which to consider new physics in gravitational theory.

The connection of TG is curvature-free \cite{Bahamonde:2021gfp,Krssak:2018ywd,Cai:2015emx} and produces an altogether novel formulation of gravity. There exists a particular combination of scalars in the theory that can produce a teleparallel equivalent of general relativity (TEGR), which is dynamically equivalent to GR but may have different IR completions \cite{Mylova:2022ljr}. A natural consequence of this is that both GR and TEGR are identical at the level of phenomenological predictions. Taking the same rationale as in $f(\lc{R})$ gravity \cite{Sotiriou:2008rp,DeFelice:2010aj,Capozziello:2011et}, TEGR can be generalized to an $f(T)$ general class of models \cite{Ferraro:2006jd,Ferraro:2008ey,Bengochea:2008gz,Linder:2010py,Chen:2010va,Bahamonde:2019zea,Paliathanasis:2017htk,Farrugia:2020fcu,Farrugia:2016qqe,Finch:2018gkh,Bahamonde:2021srr,Bahamonde:2020bbc} where the TEGR Lagrangian is simple the torsion scalar $T$. This is an interesting model since it produces organically second-order equations of motion and agrees with the polarization modes of GR despite being fundamentally different.

$f(T)$ cosmology has been probed in various scenarios. At background level, $f(T)$ models have probed against several different types of data showing consistent results with the relatively high value of the Hubble constant \cite{Briffa:2021nxg}. Most recently this has been in work that incorporates the latest ${\rm Pantheon}^{+}$ data set \cite{Briffa:2023ern,Briffa:2020qli,Cai:2019bdh,Ren:2022aeo}. There has also been working that probes the early Universe using either BBN \cite{Benetti:2020hxp}, or constraints from the CMB \cite{Nunes:2018evm,Kumar:2022nvf,Nunes:2018xbm} with interesting results related to the best-fit values of the model parameters compared with both early and late time constraints. Also, constraints from primordial black holes seem to be consistent with a wide range of observations~\cite{Papanikolaou:2022hkg} giving more freedom to the possible cosmological models. From foundational physics recent work has also been done on nonflat cosmologies~\cite{Bahamonde:2022ohm} which may open a way to compare recent proposals in the literature~\cite{DiValentino:2019qzk} on the topic in $\Lambda$CDM cosmology. Other works in the literature have explored possible effects on the fine structure constant~\cite{LeviSaid:2020mbb} which are consistent with there being no variation. 

In the current work, we are interested in determining constraints on models of $f(T)$ cosmology using $f\sigma_8(z)$ data. This window into the large-scale structure of the Universe will be a key tool in understanding the viability of new proposals for cosmology. The topic has been explored for previous data sets in Ref.~\cite{Nesseris:2013jea} where it was shown that $f(T)$ cosmological models are largely consistent with this probe of large-scale structure. Later in Ref.~\cite{Anagnostopoulos:2019miu} these analyses were combined with background data sets which gave stricter constraints on model parameters. Recently the $f\sigma_8(z)$ data has also been used to check for consistency with background data in the context of model-independent approaches to reconstructing modified cosmological models~\cite{LeviSaid:2021yat}. We start by briefly introducing the background of TG and its predictions for $f\sigma_8$ in Sec.~\ref{sec:TG_intro}. We then discuss our observational data sets under consideration in Sec.~\ref{sec:obs_data}. This is then followed by a presentation of our model constraints in Sec.~\ref{sec:results}, and a comparative analysis in Sec.~\ref{sec:analysis}. The results are summarized and discussed in our Conclusion in Sec.~\ref{sec:conclusion}.

%%%%%%%%%%%%%%%%%%%%%%%%%%%%%%%%%%%%%%%%%%%%%%%%%%%%%%%%%%
%%%%%%%%%%%%%%%%%%%%%%%%%%%%%%%%%%%%%%%%%%%%%%%%%%%%%%%%%%

\section{\texorpdfstring{$f(T)$}{} Gravity and Scalar Perturbations} \label{sec:TG_intro}

TG is based on the exchange of curvature with torsion through the geometric connection, which is realized through the teleparallel connection $\Gamma^{\sigma}_{\mu\nu}$. This renders all measures of curvature identically zero since $\lc{\Gamma}^{\sigma}_{\mu\nu}$ is curvature-less and satisfies metricity. While the regular Levi-Civita objects remain nonvanishing, such as the regular Riemann tensor $\udt{\lc{R}}{\sigma}{\mu\nu\rho} \neq 0$, its torsional analogue will vanish $\udt{R}{\sigma}{\mu\nu\rho} = 0$. This means that a different class of torsional objects must be utilized to express the geometric deformation that embodies gravitation effects from the energy-momentum tensor (see reviews in Refs. \cite{Bahamonde:2021gfp,Krssak:2018ywd,Cai:2015emx,Aldrovandi:2013wha}).

Another important aspect of TG is that it can be expressed in terms of a tetrad field that embodies the gravitational effect of the metric tensor through 
\begin{align}\label{metric_tetrad_rel}
    g_{\mu\nu}=\udt{e}{a}{\mu}\udt{e}{b}{\nu}\eta_{ab}\,,& &\eta_{ab} = \dut{E}{A}{\mu}\dut{E}{B}{\nu}g_{\mu\nu}\,,
\end{align}
where $\dut{E}{A}{\mu}$ is the tetrad inverse, and where Latin indices represent local inertial coordinates while Greek ones denote general spacetime coordinates. For internal consistency, these tetrads must also satisfy orthogonality conditions
\begin{align}
    \udt{e}{a}{\mu}\dut{E}{B}{\mu}=\delta^a_b\,,&  &\udt{e}{a}{\mu}\dut{E}{A}{\nu}=\delta^{\nu}_{\mu}\,.
\end{align}
The tetrad encompasses the gravitational freedom of the metric, while the Lorentz invariance freedom can be assigned to the spin connection $\udt{\omega}{a}{b\mu}$. The teleparallel connection can then be defined as \cite{Cai:2015emx,Krssak:2018ywd}
\begin{equation}
    \Gamma^{\sigma}_{\nu\mu}:= \dut{e}{a}{\sigma}\partial_{\mu}\udt{e}{a}{\nu} + \dut{E}{A}{\sigma}\udt{\omega}{a}{b\mu}\udt{e}{b}{\nu}\,.
\end{equation}
Given a metric, the infinite possible tetrad solutions in Eq.~\eqref{metric_tetrad_rel} are balanced by the spin connection components to retain general diffeomorphism invariance of the underlying theory. Together, the tetrad-spin connection pair defines a spacetime.

The tensor structure of TG can be written in terms of the torsion tensor which is defined as
\begin{equation}
    \udt{T}{\sigma}{\mu\nu} := -2\Gamma^{\sigma}_{[\mu\nu]}\,,
\end{equation}
where square brackets denote an antisymmetric operator which represents the translation field strength of TG \cite{Aldrovandi:2013wha}. Taking a suitable choice of contractions of the torsion tensor, we can write the torsion scalar as \cite{Hayashi:1979qx,Bahamonde:2017wwk}
\begin{equation}
    T:=\frac{1}{4}\udt{T}{\alpha}{\mu\nu}\dut{T}{\alpha}{\mu\nu} + \frac{1}{2}\udt{T}{\alpha}{\mu\nu}\udt{T}{\nu\mu}{\alpha} - \udt{T}{\alpha}{\mu\alpha}\udt{T}{\beta\mu}{\beta}\,,
\end{equation}
which is dependent only on the teleparallel connection in an analogous way to the dependence of the Riemann tensor on the Levi-Civita connection.

Naturally, the teleparallel Ricci scalar will vanish, namely $R = 0$, while the teleparallel-based Ricci scalar turns out to be equal to the torsion scalar (up to a total divergence term), that is 
\begin{equation}\label{LC_TG_conn}
    R=\lc{R} + T - B = 0\,.
\end{equation}
where $B = (2/e)\partial_\mu (eT^\mu)$ is a boundary term and $e=\det\left(\udt{e}{a}{\mu}\right)=\sqrt{-g}$ is the tetrad determinant. This relation ensures that the equations of motion of the Einstein-Hilbert action will be the same as those from the so-called teleparallel equivalent of general relativity (TEGR) which is based on the linear form of the torsion scalar. TEGR is dynamically equivalent to GR but may differ in its UV completion. The TEGR Lagrangian can readily be generalized using the same reasoning as $f(\lc{R})$ gravity \cite{DeFelice:2010aj,Capozziello:2011et}, except that we now consider an arbitrary function of the torsion scalar through \cite{Ferraro:2006jd,Ferraro:2008ey,Bengochea:2008gz,Linder:2010py,Chen:2010va,Capozziello:2022zzh,Calza:2023hhi}
\begin{equation}\label{f_T_ext_Lagran}
    \mathcal{S}_{\mathcal{F}(T)}^{} =  \frac{1}{2\kappa^2}\int \mathrm{d}^4 x\; e f(T) + \int \mathrm{d}^4 x\; e\mathcal{L}_{\text{m}} = \frac{1}{2\kappa^2}\int \mathrm{d}^4 x\; e\left(-T + \mathcal{F}(T)\right) + \int \mathrm{d}^4 x\; e\mathcal{L}_{\text{m}}\,,
\end{equation}
where $\kappa^2=8\pi G$, $\mathcal{L}_{\text{m}}$ is the matter Lagrangian, and $e=\det\left(\udt{e}{a}{\mu}\right)=\sqrt{-g}$ is the tetrad determinant.

The most impactful practical difference between $f(\lc{R})$ and $\mathcal{F}(T)$ gravity theories is that the boundary term is no longer a total divergence term for $f(\lc{R})$ gravity and so leads to fourth order equations of motion while $\mathcal{F}(T)$ gravity retains generally second order equations of motion. This may be advantageous both for numerical reasons as well as theoretical motivations when considering certain types of ghosts in the theory. By taking a variation with respect to the tetrad field, the field equations follow
\begin{align}\label{ft_FEs}
    \dut{W}{\rho}{\mu} := e^{-1} &\partial_{\nu}\left(e\dut{E}{A}{\rho}\dut{S}{\rho}{\mu\nu}\right)\left(-1 + \mathcal{F}_T\right) - \dut{E}{A}{\lambda} \udt{T}{\rho}{\nu\lambda}\dut{S}{\rho}{\nu\mu} \left(-1 + \mathcal{F}_T\right) + \frac{1}{4}\dut{E}{A}{\mu}\left(-T + \mathcal{F}(T)\right) \nonumber\\
    & + \dut{E}{A}{\rho}\dut{S}{\rho}{\mu\nu}\partial_{\nu}\left(T\right)\mathcal{F}_{TT}  + \dut{E}{B}{\lambda}\udt{\omega}{b}{a\nu}\dut{S}{\lambda}{\nu\mu}\left(-1 + \mathcal{F}_T\right) = \kappa^2 \dut{E}{A}{\rho} \dut{\Theta}{\rho}{\mu}\,,
\end{align}
where subscripts denote derivatives ($\mathcal{F}_T=\partial \mathcal{F}/\partial T$ and  $\mathcal{F}_{TT}=\partial^2 \mathcal{F}/\partial T^2$), and $\dut{\Theta}{\rho}{\nu} = \delta \mathcal{L}_{\text{m}}/ \delta \udt{e}{a}{\mu}$ is the regular energy-momentum tensor. This limits to TEGR as $\mathcal{F}(T) \rightarrow 0$. TG contains two dynamical variables namely the tetrad and spin connection which both have associated variations of the $\mathcal{F}(T)$ action. It turns out that they can both be expressed in terms of the above tetrad variation as
\begin{equation}\label{eq:FEs}
    W_{(\mu\nu)} = \kappa^2 \Theta_{\mu\nu}\,, \quad \text{and} \quad W_{[\mu\nu]} = 0\,,
\end{equation}
where the tetrad and spin connection field equations are respectively the symmetric and antisymmetric operators acted upon $W_{\mu\nu}$ \cite{Bahamonde:2021gfp}. Indeed, a particular choice of tetrad components exists in which the spin connection components vanish, called the Weitzenb\"{o}ck gauge. Here, the spin connection equations on motion are identically satisfied. More generally, the six spin connection field equations express the Lorentz invariance freedom (three translations and three rotations), while the ten tetrad field equations manifest the gravitational equations of motion. 

We can now consider a flat homogeneous and isotropic cosmology through the regular flat Friedmann--Lema\^{i}tre--Robertson--Walker (FLRW) metric
\begin{equation}\label{FLRW_metric}
     \mathrm{d}s^2=-\mathrm{d}t^2+a^2(t) \left(\mathrm{d}x^2+\mathrm{d}y^2+\mathrm{d}z^2\right)\,,
\end{equation}
from which we can identify the Hubble parameter as $H=\dot{a}/a$, and where over-dots refer to derivatives with respect to cosmic time. It can be shown that the tetrad \cite{Krssak:2015oua,Tamanini:2012hg}
 \begin{equation}
    \udt{e}{a}{\mu} = \text{diag}\left(1,\,a(t),\,a(t),\,a(t)\right)\,,
\end{equation}
is compatible with the Weitzenb\"{o}ck gauge and so the spin connection components can be set to zero \cite{Hohmann:2019nat}. By considering the tetrad field equations~\eqref{eq:FEs}, the modified Friedmann equations can be written as 
\begin{align}
    H^2 - \frac{T}{3}\mathcal{F}_T + \frac{\mathcal{F}}{6} &= \frac{\kappa^2}{3}\rho\,,\label{eq:Friedmann_1}\\
    \dot{H}\left(1 - \mathcal{F}_T - 2T\mathcal{F}_{TT}\right) &= -\frac{\kappa^2}{2} \left(\rho + p \right)\label{eq:Friedmann_2}\,,
\end{align}
where $\rho$ and $p$ denote the energy density and pressure of the matter components respectively.

TG can be used to interpret large-scale structure data by considering scalar perturbations of the flat FLRW metric. We probe this data through the growth rate measurements of $f\sigma_8^{}(z)$ from RSD. This is expressed by the logarithm derivative of the matter perturbation $\delta(z)$ with respect to the logarithm of the cosmic scale factor, namely
\begin{equation}\label{eq:fz_growth}
    f(z)=\frac{\mathrm{d}\ln\delta(z)}{\mathrm{d}\ln a}=-(1+z)\frac{\mathrm{d}\ln\delta(z)}{\mathrm{d}z}=-(1+z)\frac{\delta^\prime(z)}{\delta(z)}\,,
\end{equation}
where a prime denotes a derivative with respect to redshift $z=a^{-1}-1$. On the other hand, the linear theory root--mean--square mass fluctuation within a sphere of radius $8h^{-1}$ Mpc can be expressed as
\begin{equation}\label{eq:s8}
    \sigma_8^{}(z)=\sigma_{8,0}^{}\frac{\delta(z)}{\delta_0^{}}\,,
\end{equation}
where a 0--subscript denotes the respective value at $z=0$. Thus, the growth rate can be generally written as
\begin{equation}\label{eq:fs8}
    f\sigma_8^{}(z)=-(1+z)\frac{\sigma_{8,0}^{}}{\delta_0^{}}\delta^\prime(z)\,,
\end{equation}
which directly leads to the normalized form of $\delta^\prime(z)$ through
\begin{equation}\label{eq:delta_prime}
    \frac{\delta^\prime(z)}{\delta_0^{}}=-\frac{1}{\sigma_{8,0}^{}}\frac{f\sigma_8^{}(z)}{(1+z)}\,.
\end{equation}
By integrating this expression, the normalized redshift evolution of the matter perturbation can be written as
\begin{equation}\label{eq:delta}
    \frac{\delta(z)}{\delta_0^{}}=1-\frac{1}{\sigma_{8,0}}\int_0^z\frac{f\sigma_8^{}(\tilde{z})}{(1+\tilde{z})}\,\mathrm{d}\tilde{z}\,,
\end{equation}
while Eq. (\ref{eq:delta_prime}) also gives the second-order derivative 
\begin{equation}\label{eq:delta_prime_prime}
    \frac{\delta^{\prime\prime}(z)}{\delta_0^{}}=-\frac{1}{\sigma_{8,0}^{}}\left[\frac{(1+z)\,f\sigma_8^\prime(z)-f\sigma_8^{}(z)}{(1+z)^2}\right]\,.
\end{equation}
The result is the redshift evolution of $f$ can now be determined
\begin{equation}\label{eq:f_prime}
    f^\prime(z)=-\left(\frac{\delta^\prime(z)}{\delta_0^{}}\right)\left(\frac{\delta_0^{}}{\delta(z)}\right)-(1+z)\left[\left(\frac{\delta^{\prime\prime}(z)}{\delta_0^{}}\right)\left(\frac{\delta_0^{}}{\delta(z)}\right)-\left(\frac{\delta^\prime(z)}{\delta_0^{}}\right)^2\left(\frac{\delta_0^{}}{\delta(z)}\right)^2\right]\,.
\end{equation}

In the subhorizon limit, the linear matter perturbations equation can then be written as
\begin{equation}\label{eq:delta_dot}
    \ddot{\delta}+2H\dot{\delta}=4\pi G_\mathrm{eff}\,\rho\,\delta\,,
\end{equation}
where $G_\mathrm{eff}$ is the effective Newton's constant which is in general a function of $z$ and cosmic wave vector $\textbf{k}$ \cite{amendola2010,Hernandez:2016xci}. However for this limit and the data sets under consideration, the $G_\mathrm{eff}$ can be taken to be independent of $\textbf{k}$ . In this setting, Eq.~(\ref{eq:delta_prime2}) takes the form
\begin{equation}\label{eq:delta_prime2}
    \delta^{\prime\prime}(z)+\left(\frac{H^\prime(z)}{H(z)}-\frac{1}{1+z}\right)\delta^\prime(z)=\frac{3}{2}\frac{G_\mathrm{eff}(z)}{G_N^{}}\left(\frac{H_0^{}}{H(z)}\right)^2\Omega_{m,0}^{}\,(1+z)\,\delta(z)\,,
\end{equation}
where $\Omega_{m,0}^{}$ denotes the current matter fractional density, $H_0^{}$ is Hubble's constant, and $G_N^{}$ is Newton's gravitational constant. We can also express this relation in terms of the growth rate $f$, giving
\begin{equation}\label{eq:growth_rate}
    f^2(z)+\left[2-(1+z)\frac{H^\prime(z)}{H(z)}\right]f(z)-(1+z)f^\prime(z)=\frac{3}{2} \frac{G_\mathrm{eff}^{}(z)}{G_N^{}} \left(\frac{H_0}{H(z)}\right)^2\Omega_{m,0}^{}(1+z)^3\,.
\end{equation}
In the case of $\mathcal{F}(T)$ gravity, the linear matter perturbation evolution Eq.~\eqref{eq:delta} is expressed through \cite{Nunes:2018xbm,Golovnev:2018wbh,LeviSaid:2020mbb,Sahlu:2019bug}
\begin{equation}\label{eq:Qz}
    G_\mathrm{eff}^{}(z)=\frac{G_N^{}}{1+\mathcal{F}_T^{}(z)}\,.
\end{equation}
Hence, the linear matter perturbation equation is sensitive to $\mathcal{F}(T)$ gravity and so large-scale structure data sets can be used to probe observational constraints on potential models.

In order to utilize equations like Eq.~\eqref{eq:delta_prime2} and Eq.~\eqref{eq:growth_rate}, we need to determine the value of $H'(z)$, obtained from the second Friedmann equation, Eq.~\eqref{eq:Friedmann_2}, via
\begin{equation}\label{eq:H_prime}
    H'(z) = \frac{\kappa^2 (p + \rho)}{2 H (1+z) \left(1 - \mathcal{F}_T - 2T\mathcal{F}_{TT}\right)}.
\end{equation}

%%%%%%%%%%%%%%%%%%%%%%%%%%%%%%%%%%%%%%%%%%%%%%%%%%%%%%%%%%
%%%%%%%%%%%%%%%%%%%%%%%%%%%%%%%%%%%%%%%%%%%%%%%%%%%%%%%%%%

\section{Observational data} \label{sec:obs_data}

In this section, we will provide an overview of how we utilized observational data in our research to evaluate the most favorable $f(T)$ models from the literature. We include a variety of observational data, such as the Pantheon+ sample of SNIa data, Baryon acoustic oscillations (BAO) data, Cosmic Chronometers (CC) data, and growth rate (redshift space distortion, RSD) data. To analyze the data, we implemented an MCMC (Monte Carlo Markov Chain) approach using the \textit{emcee} package publicly available at Ref.~\cite{2013PASP..125..306F}.  This enabled us to constrain the model and cosmological parameters and thus, investigate the posterior of the parameter space. This yielded one-dimensional distributions that illustrate the parameter's posterior distribution, and two-dimensional distributions that demonstrate the covariance between two different parameters. These distributions were complemented by the 1- and 2-$\sigma$ confidence levels as will be shown in Sec.~\ref{sec:results}.

%%%%%%%%%%%%%%%%%%%%%%%%%%%%%%%%%%%%%%%%%%%%%%%%%%%%%%%%%%
%%%%%%%%%%%%%%%%%%%%%%%%%%%%%%%%%%%%%%%%%%%%%%%%%%%%%%%%%%

\subsection{Cosmic Chronometers}

Cosmic Chronometers (CC) offer a useful tool to directly constrain the Hubble rate $H(z)$ at different redshifts. To this end, we adopt thirty-one data points as compiled from Ref.~\cite{2014RAA....14.1221Z,Jimenez:2003iv,Moresco:2016mzx,Simon:2004tf,2012JCAP...08..006M,2010JCAP...02..008S,Moresco:2015cya}. The CC method involves using spectroscopic dating techniques on passively-evolving galaxies to estimate the age difference between two galaxies at different redshifts. By measuring this age difference, $\Delta z / \Delta t$ can be inferred from observations, allowing for the computation of $H(z) = -(1+z)^{-1} \Delta z/ \Delta t$. Thus, CC data were found to be particularly reliable because they are independent of any cosmological model, avoid any complex integration, and rely on the absolute age determination of galaxies \cite{Jimenez:2001gg}.

In our MCMC analysis, we used $\chi^2_\mathrm{CC}$ to evaluate the agreement between the theoretical Hubble parameter values $H(z_i,\Theta)$, with model parameters $\Theta$, and the observational Hubble data values $H_{\mathrm{obs}}(z_i)$, with an observational error of $\sigma_H(z_i)$. The $\chi^2_\mathrm{CC}$ was calculated using 
\begin{equation}
    \chi^2_\mathrm{CC} = \sum^{39}_{i=1} \frac{\left(H(z_i,\Theta) - H_{\mathrm{obs}}(z_i)\right)^2}{\sigma^2_H(z_i)} \,.
\end{equation}

%%%%%%%%%%%%%%%%%%%%%%%%%%%%%%%%%%%%%%%%%%%%%%%%%%%%%%%%%%
%%%%%%%%%%%%%%%%%%%%%%%%%%%%%%%%%%%%%%%%%%%%%%%%%%%%%%%%%%

\subsection{Type Ia Supernovae - Pantheon+ data set}

The SNIa dataset used in this study is the Pantheon+ ($\mathrm{PN}^+$\,\&\,SH0ES) sample \cite{Scolnic:2021amr}, which is one of the largest compilations of SNIa data and contains 1701 SNIa measurements spanning a redshift range of 0.0 to 2.26. The Pantheon+ analysis (\cite{Brout:2022vxf}), incorporates SH0ES Cepheid host distance anchors, \cite{Riess:2019cxk} in the likelihood, effectively addressing the degeneracy between parameters $M$ and $H_0$ in Type Ia supernova analysis. Following this approach, we also adopt the label `PN+ \& SH0ES' to signify our incorporation of the same methodology.

SNIa can be used to determine the expansion rate of the universe, $H(z)$, by measuring the observed apparent magnitude, $m$, and the absolute magnitude, $M$, through the equation:

\begin{equation} \label{eq:dist_mod}
    \mu(z_i, \Theta) = m - M = 5 \log_{10}[D_L(z_i, \Theta)] + 25 \,,
\end{equation}
where  $\Theta$ represents the set of cosmological parameters that describe the universe, $z_i$ is the redshift of the SNIa measurement and $D_L(z_i, \Theta)$ is the luminosity distance given by 
\begin{equation}
    D_L(z_i, \Theta) = c(1+z_i) \int_0^{z_i} \frac{dz'}{H(z', \Theta)} \,, 
\end{equation}
where $c$ is the speed of light. To calibrate the observed apparent magnitude of each SNIa, a fiducial absolute magnitude $M$ is used, as shown in Eq.~\eqref{eq:dist_mod}. In our MCMC analyses, we treat $M$ as a nuisance parameter.

To constrain the cosmological parameters, the minimum $\chi^2$ is calculated through \cite{SNLS:2011lii}, 
\begin{equation}
    \chi^2_{\mathrm{SN}} = (\Delta \mu(z_i), \Theta))^T C^{-1}_\mathrm{SN} (\Delta \mu(z_i), \Theta)) \,,
\end{equation}
where $(\Delta \mu(z_i), \Theta)) = (\mu(z_i), \Theta) - \mu(z_i)_{\mathrm{obs}}$ and $C_\mathrm{SN}$ is the corresponding covariance matrix which accounts for the statistical and systematic uncertainties. 

%%%%%%%%%%%%%%%%%%%%%%%%%%%%%%%%%%%%%%%%%%%%%%%%%%%%%%%%%%
%%%%%%%%%%%%%%%%%%%%%%%%%%%%%%%%%%%%%%%%%%%%%%%%%%%%%%%%%%

\subsection{Baryon Acoustic oscillations (BAO)}

In this study, we use a variety of observational missions to constrain cosmological parameters. These data sets include measurements from the 6dF Galaxy Survey (6dFGS) at an effective redshift of $z_{\mathrm{eff}} = 0.106$ \cite{2011MNRAS.416.3017B}, the BOSS Data Release 11 (DR11) quasar Lyman-alpha measurements at an effective redshift of $z_{\mathrm{eff}} = 2.4$ \cite{Bourboux:2017cbm}, and the SDSS Main Galaxy sample at an effective redshift of $z_{\mathrm{eff}} = 0.15$ \cite{Ross:2014qpa}. Additionally, measurements of the Hubble parameter and the corresponding comoving angular diameter at $z_{\mathrm{eff}} = {0.38,0.51}$ were obtained from the third generation of the SDSS mission (SDSS BOSS DR12) \cite{eBOSS:2020yzd}. We also include $H(z)$ measurements and angular diameter distances at $z_{\mathrm{eff}} = {0.98,1.23,1.52,1.94}$ from the fourth generation of the SDSS mission (SDSS-IV BOSS DR12) \cite{Zhao:2018gvb}.

As has been mentioned these different data sets report different observational quantities that are related to one another. For the BAO data sets under consideration we compute the Hubble distance $D_H(z)$ given by $D_H(z) = \frac{c}{H(z)}$. We also consider the angular diameter distance $D_A(z)$ defined as 
\begin{equation}
    D_A(z) = \frac{c}{1+z} \int^z_0 \frac{dz'}{H(z')} \,,
\end{equation}
from which two other quantities can be derived. The first is the comoving angular diameter distance $D_M$ given trough $D_M = (1+z)D_A(z)$ whilst the second one is the volume average-distance given by
\begin{equation}
    D_V(z) = (1+z)^2 \left[D_A(z)^2 \frac{cz}{H(z)}\right]^{\frac{1}{3}} \,. 
\end{equation}

Using the BAO results, we calculate the corresponding combination of results $\mathcal{G}(z_i)=D_V(z_i)/r_s(z_d),\allowbreak\,r_s(z_d)/D_V(z_i),\allowbreak\,D_H(z_i),\allowbreak\,D_M(z_i)(r_{s,\mathrm{fid}}(z_d)/r_s(z_d)),\allowbreak\,H(z_i)(r_s(z_d)/r_{s,\mathrm{fid}}(z_d)),\allowbreak\,D_A(z_i)(r_{s,\mathrm{fid}}(z_d)/r_s(z_d))$. In this case, we require the comoving sound horizon at the end of the baryon drag epoch at $z_d \sim 1059.94$ \cite{Planck:2018vyg} which can be calculated using 
\begin{equation}
    r_s(z)=\int_z^\infty\frac{c_s(\tilde{z})}{H(\tilde{z})}\,\mathrm{d}z=\frac{1}{\sqrt{3}}\int_0^{1/(1+z)}\frac{\mathrm{d}a}{a^2H(a)\sqrt{1+\left[3\Omega_{b,0}/(4\Omega_{\gamma,0})\right]a}}\,,
\end{equation}
where we have adopted a fiducial value of $r_{s,\mathrm{fid}}(z_d)=147.78,\mathrm{Mpc}$ \cite{Planck:2018vyg} with an assumption of $\Omega_{b,0}=0.02242$ \cite{Planck:2018vyg} and $T_{0}=2.7255,\mathrm{K}$ \cite{2009ApJ...707..916F}
Therefore, the corresponding $\chi^2$ is calculated through
\begin{equation}
    \chi^2_{\text{BAO}}(\Theta) = \Delta G(z_i,\Theta)^T C_{\text{BAO}}^{-1}\Delta G(z_i,\Theta)
\end{equation}
where $\Delta G(z_i,\Theta) = G(z_i,\Theta)-G_{\text{obs}}(z_i)$and $C_{\text{BAO}}$ is the corresponding covariance matrix for the BAO observations. The total $\chi^2_{\mathrm{BAO}}$ is therefore the sum of all the individual $\chi^2$ corresponding to each data set.

%%%%%%%%%%%%%%%%%%%%%%%%%%%%%%%%%%%%%%%%%%%%%%%%%%%%%%%%%%
%%%%%%%%%%%%%%%%%%%%%%%%%%%%%%%%%%%%%%%%%%%%%%%%%%%%%%%%%%

\subsection{Growth rate data (RSD)}

The final data compilation we used in this work is the growth rate data compilation presented in Table~II of Appendix~A in Ref~\cite{Alestas:2022gcg}. This dataset consists of measurements of the growth of cosmic structure and is commonly referred to as Redshift Space Distortion (RSD) data, due to a phenomenon that occurs during observations at both large and small scales. The peculiar velocity of galaxies causes high-density regions of the universe to appear elongated in the line-of-sight direction at small scales and flattened at large scales. As a result, maps of galaxies where distances are measured from spectroscopic redshifts exhibit anisotropic deviations from the true galaxy distribution. Therefore, RSD measurements can provide valuable insights into the large-scale structure of the Universe, which is shaped by the underlying theory of gravity governing the evolution and formation of cosmic structure. Consequently, RSD presents a promising approach to testing modified theories of gravity.

The growth rate $f(z)$, can be estimated using RSD cosmological probes as a way to constrain cosmological models \cite{Lambiase:2018ows, Gonzalez:2016lur, Gupta:2011kw}. However, instead of reporting the growth rate directly, LSS surveys typically report the density-weighted growth rate, $f\sigma_8 \equiv f(z) \sigma_8(z)$ which is bias-independent \cite{Kazantzidis:2018rnb}. Furthermore, we should acknowledge the Alcock-Paczynski (AP) effect, which emerges from the requirement of assuming a cosmological model for the conversion of redshifts into distances. However, recent studies, including those outlined in Ref.~\cite{LeviSaid:2021yat}, have indicated that the influence of this effect is minimal and thus has been disregarded in our analysis.

Therefore, while the theoretical prediction can be made from Eq.~\eqref{eq:fs8}, the aforementioned RSD data can be used to constrain cosmological parameters, in particular $\sigma_{8,0}$ such the corresponding $\chi^2_{\mathrm{min}}$ can be given as
\begin{equation}
    \chi^2_\mathrm{RSD}  = \Delta Q(z_i,\Theta)^T C_{\text{RSD}}^{-1}\Delta Q(z_i,\Theta)
\end{equation}
where $Q(z_i,\Theta) =(f\sigma_8(z_i,\Theta)_{\mathrm{theo}} - f\sigma{_8{}_\mathrm{obs}}(z_i))$ and $C_{\text{RSD}}^{-1}$ is the inverse covariance matrix which is assumed to be a diagonal matrix except for the WiggleZ subset data which can be found in Refs.~\cite{Blake:2012pj}. Therefore the total covariance matrix can be written as 
\begin{equation}
C_{\text{RSD}} = 
\begin{pmatrix}
    \sigma_1^2 & 0 & 0 & \dots\\
    0 & C_{\text{WiggleZ}} & 0 & \dots \\
    0 & 0 & \dots & \sigma_N^2 \\
\end{pmatrix} \,.
\end{equation}

To compute $Q(z_i,\Theta)$, we start by obtaining the theoretical values of $f\sigma_8(z_i,\Theta)_{\mathrm{theo}}$ using Eq.~\eqref{eq:delta_prime2} along with Eq.~\eqref{eq:fs8}. On the other hand, $f\sigma_8(z_i,\Theta)_{\mathrm{obs}}$ is extracted from the compiled observational data as described earlier.

%%%%%%%%%%%%%%%%%%%%%%%%%%%%%%%%%%%%%%%%%%%%%%%%%%%%%%%%%%
%%%%%%%%%%%%%%%%%%%%%%%%%%%%%%%%%%%%%%%%%%%%%%%%%%%%%%%%%%

\section{Results} \label{sec:results}

In this section, we delve into the outcomes yielded by various combinations of data sets. Each subsection is dedicated to distinct $f(T)$ models, selected based on their potential to accurately reflect our cosmological history, as demonstrated in previous studies \cite{dosSantos:2021owt,Nunes:2016qyp,Briffa:2023ern,Briffa:2021nxg,Sandoval-Orozco:2023pit}. For every model, we showcase the 1$\sigma$ and 2$\sigma$ uncertainty ranges. These results are accompanied by a table showing the final results that include the Hubble constant ($H_0$) expressed in ${\rm km s}^{-1} {\rm Mpc}^{-1}$, the matter density parameter $\Omega_{m,0}$ and $\sigma_{8,0}$ along with other model parameters. This will give us an insight into how the different models and data set combinations affect the $H_0$ tension and $S_{8,0}$ cosmological tensions.

\subsection{Power Law Model}

The power-law model, initially proposed by Bengochea and Ferraro in their work \cite{Bengochea:2008gz}, was adopted due to its capability to replicate the observed late-universe acceleration without the need for dark energy. Henceforth, this model will be referred to as $f_1$CDM  and is specified by
\begin{equation} \label{eq:PLM}
    \mathcal{F}_1(T) = \alpha_1(-T)^{p_1} \,.
\end{equation}
This power-law form is characterized by two constants, namely $\alpha_1$ and $p_1$ which is introduced as a modification to $\mathcal{F}(T)$ function. Being a constant $\alpha_1$ can be evaluated at any time and therefore, it can be calculated at the current time by using the Friedmann equation~\eqref{eq:Friedmann_1}, which results in
\begin{equation} \label{eq:alpha1}
    \alpha_1 = (6H_0^2)^{1-p_1} \frac{1- \Omega_{m,0} - \Omega_{r,0}}{1-2p_1} \,,
\end{equation}

\begin{table}
\resizebox{\textwidth}{!}{%
\centering  
 %   \hspace{-2cm}
    \begin{tabular}{cccccc}
        \hline
		Data sets & $H_0 \mathrm{\hspace{0.15cm}[km \hspace{0.1cm} s^{-1} \hspace{0.1cm}Mpc ^{-1}]}$ & $\Omega_{m,0}$ & $p_1$ & $\sigma_{8,0}$ & $M$ \\ 
		\hline
		CC + BAO  & $68.1^{+1.2}_{-1.4}$ & $0.314^{+0.034}_{-0.033}$ & $-0.09^{+0.24}_{-0.30}$ & -- & -- \\ 
		CC +  BAO + RSD & $69.17\pm 0.81$ & $0.287^{+0.016}_{-0.020}$ & $-0.09^{+0.17}_{-0.20}$ & $0.785\pm 0.035$ & -- \\ 
		$\mathrm{PN}^+$\,\&\,SH0ES  + RSD & $73.7\pm 1.0$ & $0.290^{+0.019}_{-0.018}$ & $0.076^{+0.075}_{-0.102}$ & $0.817^{+0.037}_{-0.035}$ & $-19.252^{+0.029}_{-0.030}$ \\ 
		CC + $\mathrm{PN}^+$\,\&\,SH0ES  + BAO  & $69.45^{+0.69}_{-0.58}$ & $0.316^{+0.028}_{-0.029}$ & $-0.06^{+0.19}_{-0.22}$ & -- & $-19.375\pm 0.017$ \\ 
		CC + $\mathrm{PN}^+$\,\&\,SH0ES  + BAO + RSD & $69.90\pm 0.58$ & $0.289^{+0.016}_{-0.018}$ & $0.014^{+0.091}_{-0.125}$ & $0.810^{+0.036}_{-0.033}$ & $-19.367^{+0.016}_{-0.017}$ \\ 
		\hline
    \end{tabular}
    }
     \caption{Exact results for $f_1$ model that include the parameters $H_0$, $\Omega_{m,0}$ and $p_1$. The $\sigma_{8,0}$ parameter and the nuisance parameter $M$, are provided for data sets that include RSD or $\mathrm{PN}^+$\,\&\,SH0ES, respectively  otherwise, they are left empty.}
    \label{tab:PLM}
\end{table}
where the density parameter for matter and radiation is represented by $\Omega_{m,0}$ and $\Omega_{r,0}$ respectively. Consequently, by utilizing Eq.~\eqref{eq:alpha1}, the $f_1$CDM model introduces only one new parameter, namely $p_1$, thereby enhancing its elegance and simplicity. In combination with the parameters $H_0$, $\Omega_{m,0}$, and $\sigma_{8,0}$, we determine the model parameter $p_1$ through MCMC analyses performed on observational data. 

By substituting the above equation into the Friedmann equation~\eqref{eq:Friedmann_1}, we derive the resulting Friedmann equation for this model
\begin{equation}\label{eq:PLM_FE}
    E^2(z) = \Omega_{m,0}(1+z)^3 + \Omega_{r,0}(1+z)^4 + (1-\Omega_{m,0} - \Omega_{r,0}) E^{2p_1}(z) ,,
\end{equation}
where $E(z) := \frac{H(z)}{H_0}$. It is worth noting that this equation cannot be solved analytically, so we employ numerical methods to calculate $E(z)$ at each redshift point. Consequently, using MCMC analysis, we solve for each redshift point where observational data exists to obtain parameter values. Specifically, we extract values for $H_0$, $\Omega_{m,0}$, and $p_1$. Additionally, when RSD data is included, we can obtain values for $\sigma_{8,0}$ allowing us to assess the model's compatibility with observational data and refine our understanding of the Universe's properties.b When examining Eq.~\eqref{eq:PLM}, it becomes apparent that the $\Lambda$CDM limit can be attained when $p_1 = 0$. On the other hand, when $p_1 = 1$ the model converges to the GR limit. In this scenario, an additional component in the Friedmann equation emerges, introducing a re-scaled gravitational constant term within the density parameters.

\begin{figure}[]
    \centering
    \includegraphics[width = 0.8\textwidth]{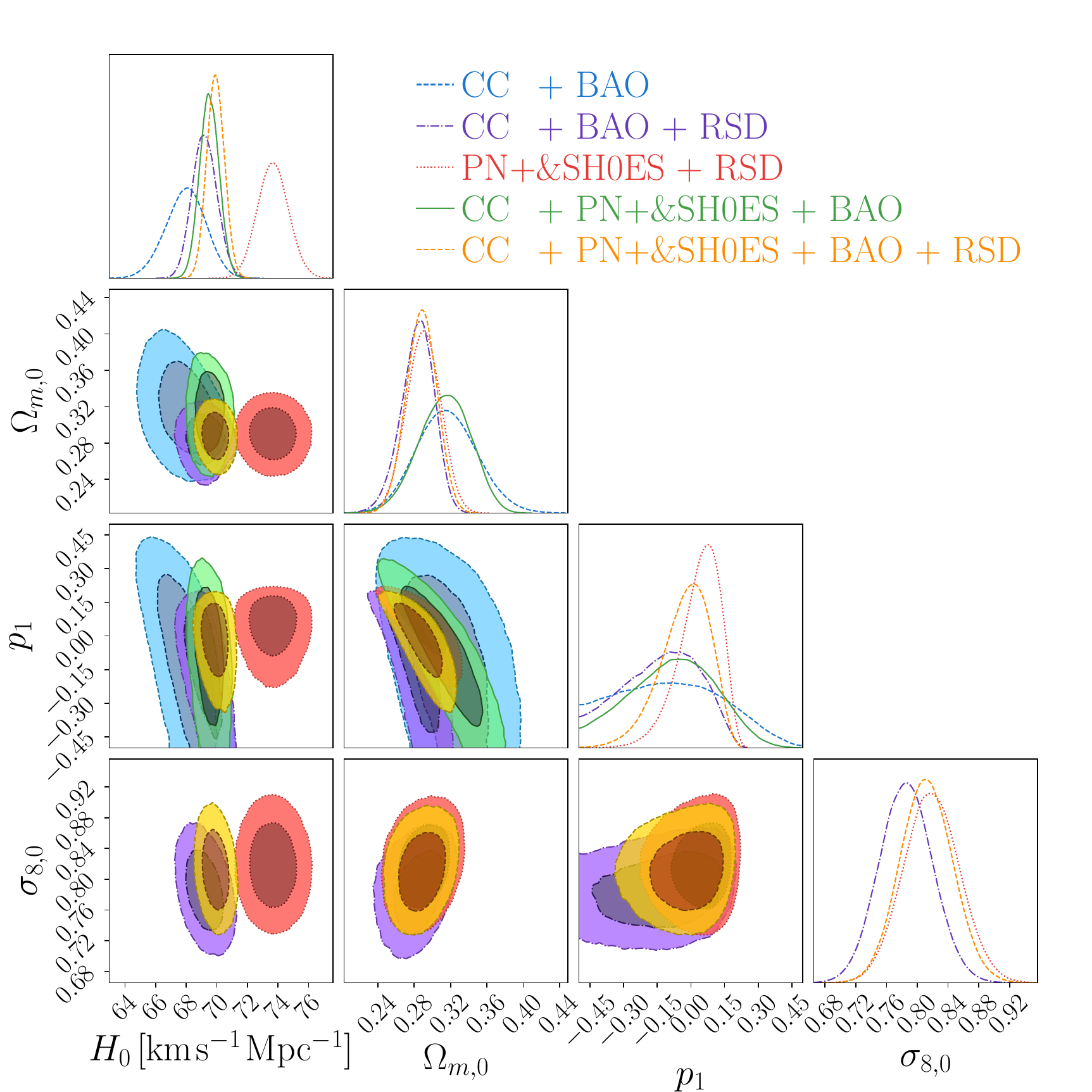}
    \caption{Confidence contours and posterior distributions for the $f_1$CDM model (Power Law Model) parameters, including $H_0$, $\Omega_{m,0}$, and $p_1$. In cases where the RSD data is incorporated (purple, red, and yellow contours), the $\sigma_{8,0}$ parameter is also displayed.}
    \label{fig:PLM}
\end{figure}

The contour plots together with their posteriors for the $f_1$CDM model are shown in Fig.~\ref{fig:PLM}. In this and subsequent models, the blue and green contours represent the CC+BAO and CC+$\mathrm{PN}^++$BAO data sets, respectively. The remaining colors correspond to various data sets that incorporate the RSD data. On the other hand, the precise values for the parameters are tabulated in Table~\ref{tab:PLM}. Notably, this table highlights that the highest $H_0$ value is obtained when combining the $\mathrm{PN}^+$\,\&\,SH0ES data with the RSD data is used. This result aligns with expectations, as the $\mathrm{PN}^+$\,\&\,SH0ES data converges with values reported by the SH0ES team, which reports a value of $H_0 = 73.30 \pm 1.04 {\rm\, km \, s}^{-1} {\rm Mpc}^{-1}$ \cite{Scolnic:2021amr}. 
Conversely, the lowest value for the $H_0$ parameter is derived from the CC+BAO data sets, primarily influenced by the BAO data, which relates to conditions in the early Universe.

It is noteworthy to notice the relationship between the Hubble parameter ($H_0$) and the matter density parameter ($\Omega_{m,0}$). As $H_0$ increases, indicating a faster expansion rate of the Universe, the matter density decreases. This effect is particularly pronounced in the $\mathrm{PN}^+$\,\&\,SH0ES+RSD combination, where the Universe's energy predominantly manifests as an effective form of dark energy, driven by the elevated $H_0$ parameter.  The presence of this phenomenon is further confirmed by the apparent anti-correlation depicted in Fig.~\ref{fig:PLM}. 

As noted, the $\Lambda$CDM limit is reached for when $p_1 = 0$. Indeed, the resultant values for $p_1$ are in proximity of 0, and this limit falls within the 1$\sigma$ range. Furthermore, it can be observed that the inclusion of the RSD data leads to tighter constraints on the $p_1$ parameter. However, this effect is not exclusive to the $p_1$ parameter alone; it also extends to other parameters, as evident from the contour plots in Fig.~\ref{fig:PLM}.

Incorporating the RSD data has been instrumental in constraining the amplitude of mass fluctuations, $\sigma_{8,0}$. Upon initial examination of Table~\ref{tab:PLM}, it appears that there is a correlation between the $H_0$ parameter and $\sigma_{8,0}$, where a higher $H_0$ tends to correspond to a slightly higher $\sigma_{8,0}$. However, the contour plots suggest a more complex and degenerate relationship between these parameters. It is worth noting that the RSD data appears to have more Gaussian errors when compared to the other data sets. The influence of these growth structure data is encapsulated by the parameter $G_{\mathrm{eff}}$, which, in this model, is expressed as 
\begin{equation}
    G_{\mathrm{eff}} = \frac{G_N}{1 - \alpha_1 p_1 (-T)^{p_1-1}} \,,
\end{equation}
where the specific values for the parameters are extracted from the relevant table, and as a result, we observe that $G_{\mathrm{eff}}$ approximates $G_N$ under these conditions.

\begin{figure}[htbp]
  \begin{minipage}{0.5\textwidth}
  \captionsetup{width=0.9\textwidth}
    \includegraphics[width = 0.8\textwidth]{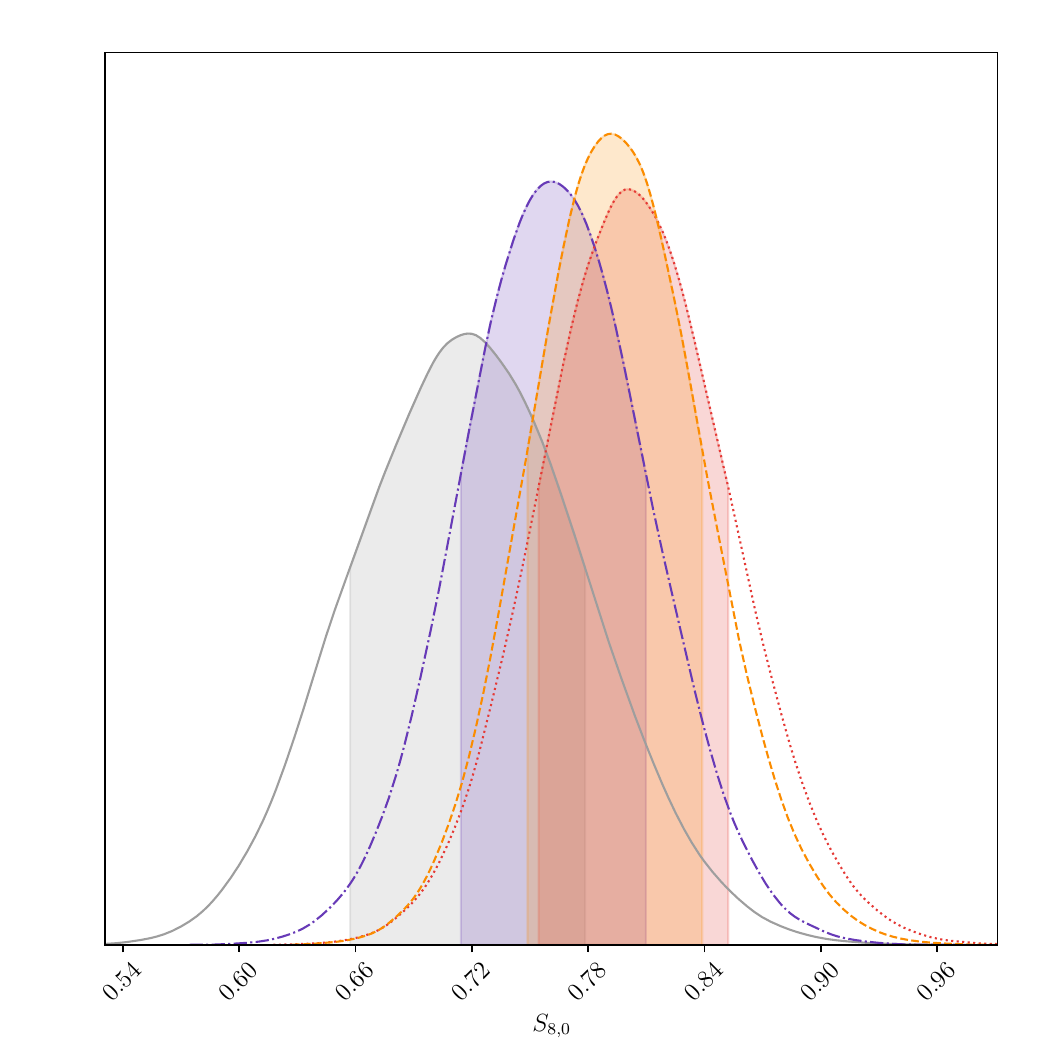}
    \captionof{figure}{Posterior distribution for the $S_{8,0}$ parameter in the $f_1$CDM model. Legend: Grey denotes the RSD data, purple corresponds to CC+BAO+RSD data, red represents the $\mathrm{PN}^+$\,\&,SH0ES + RSD dataset, while orange indicates CC + $\mathrm{PN}^+$\,\&\,SH0ES + BAO + RSD data.}
    \label{fig:PLM_S80}
  \end{minipage}
  \begin{minipage}{0.5\textwidth}
  \captionsetup{width=0.9\textwidth}
    \begin{tabular}{lc}
        \hline
		Data sets & $S_{8,0}$ \\ 
		\hline
  
        \textcolor{mygray}{\tikz[baseline=-0.75ex]\draw [thick,solid] (0,0) -- (0.5,0); RSD} & $0.718^{+0.061}_{-0.060}$ \\ 
	\texttransparent{0.7}{\textcolor{patriarch}{\tikz[baseline=-0.75ex]\draw [thick,dash dot dot] (0,0) -- (0.5,0);CC +  BAO + RSD}} & $0.761^{+0.049}_{-0.046}$ \\ 
		\texttransparent{0.7}{\textcolor{myred}{\tikz[baseline=-0.75ex]\draw [thick,dotted] (0,0) -- (0.5,0);$\mathrm{PN}^+$\,\&\,SH0ES + RSD}} & $0.801^{+0.052}_{-0.046}$ \\ 
		\texttransparent{0.8}{\textcolor{myorange}{\tikz[baseline=-0.75ex]\draw [thick,dashed] (0,0) -- (0.5,0);CC + $\mathrm{PN}^+$\,\&\,SH0ES + BAO + RSD}} & $0.792^{+0.047}_{-0.043}$ \\ 
		\hline
    \end{tabular}
    \captionof{table}{Exact $S_{8,0}$ values corresponding to various data sets for the $f_1$CDM model.}
    \label{tab:PLM_S80}
  \end{minipage}
\end{figure}

Additionally, the constraint on these parameters has enabled us to explore a tension quantified in \textbf{terms of $S_{8,0} \equiv \sigma_{8,0}\sqrt{\Omega_{m,0}/0.3}$ \cite{Preston:2023uup,Benisty:2020kdt,Brieden:2022heh, Rubira:2022xhb,Clark:2021hlo,Anchordoqui:2021gji,BeltranJimenez:2021wbq}}. The results and the posteriors of these parameters can be found in Table~\ref{tab:PLM_S80} and Fig.~\ref{fig:PLM_S80}, respectively. Reflecting the values obtained for $\sigma_{8,0}$, the highest value observed for $S_{8,0}$ was attained for the $\mathrm{PN}^+$\,\&\,SH0ES+RSD combination, measuring at $S_{8,0} = 0.801^{+0.052}_{-0.046}$. However, it is worth noting that we also provide values for the RSD data set alone to isolate the impact of RSD data on this parameter, where in this case, the value for the RSD data set reaches a minimum. Furthermore, it is evident that across all data sets, the constraints for such parameters are notably tight. We also examine the relationship between the $S_{8,0}$ parameter and the model parameter $p_1$, which is depicted in Fig.~\ref{fig:PLM_p_vs_S80} found in Appendix~\ref{sec:App_pvsS}. A significant anti-correlation exists between these two parameters. In return, this might suggest that changes in the power-law exponent ($p_1$) might have a direct impact on the amplitude of mass fluctuations ($S_{8,0}$).

%%%%%%%%%%%%%%%%%%%%%%%%%%%%%%%%%%%%%%%%%%%%%%%%%
%%%%%%%%%%%%%%%%%%%%%%%%%%%%%%%%%%%%%%%%%%%%%%%%%

\subsection{Linder Model}

The second model under consideration is the Linder Model \cite{Linder:2009jz}, explicitly developed to explain the Universe's late-time acceleration without invoking the presence of dark energy. This model introduces an exponential component that incorporates the torsion scalar, denoted as $T$, and is expressed as follows
\begin{equation}
    \mathcal{F}_2 = \alpha_2 T_0 \left(1- \mathrm{Exp}\left[-p_2 \sqrt{T/T_0} \right] \right) \,,
\end{equation}
where both $\alpha_2$ and $p_2$ are constants, while $T_0 = T|_{t=t_0}= -6H_0^2$ represents the torsion scalar at the current epoch. Similar to the previous model, if $\alpha_2$ is a constant, it can be immediately determined from the Friedman equation at the present time, yielding

\begin{equation}\label{eq:LM_alpha2}
    \alpha_2 = \frac{1- \Omega_{m,0} - \Omega_{r,0}}{(1+p_2) e^{-p_2} - 1} \,.
\end{equation}

Therefore, in the Linder Model, hereafter referred to as $f_2$CDM, the only new model parameter is denoted as $p_2$, and it will be constrained through the MCMC analysis. Consequently, Eq.~\eqref{eq:LM_alpha2} allows us to express the Friedmann Equation in terms of $H_0$, $\Omega_{m,0}$, and $p_2$, resulting in the following form
\begin{equation}
    E^2\left(z\right) = \Omega_{m,0} \left(1+z\right)^3 + \Omega_{r_0}\left(1+z\right)^4 + \frac{1 - \Omega_{m,0} - \Omega_{r_0}}{(p_2 + 1)e^{-p_2} - 1} \left[\left(1 + p_2 E(z)\right) \text{Exp}\left[-p_2 E(z)\right] - 1\right] \,.
\end{equation}

\begin{table}
\resizebox{\textwidth}{!}{%
\centering  
 %   \hspace{-2cm}
 
    \begin{tabular}{cccccc}
        \hline
		Data Sets & $H_0 \mathrm{\hspace{0.15cm}[km \hspace{0.1cm} s^{-1} \hspace{0.1cm}Mpc ^{-1}]}$ & $\Omega_{m,0}$ & $\frac{1}{p_2}$ & $\sigma_{8,0}$ & M \\ 
		\hline
		CC + BAO  & $67.2^{+1.2}_{-1.6}$ & $0.302^{+0.035}_{-0.030}$ & $0.00^{+0.37}_{-0.00}$ & -- & -- \\ 
		CC +  BAO + RSD & $66.5^{+2.2}_{-1.3}$ & $0.286^{+0.016}_{-0.015}$ & $0.39^{+0.21}_{-0.22}$ & $0.784^{+0.038}_{-0.032}$ & -- \\ 
		$\mathrm{PN}^+$\,\&\,SH0ES + RSD & $73.2\pm 1.0$ & $0.287\pm 0.013$ & $0.359^{+0.077}_{-0.071}$ & $0.770^{+0.042}_{-0.039}$ & $-19.26^{+0.33}_{-0.31}$ \\ 
		CC + $\mathrm{PN}^+$\,\&\,SH0ES + BAO  & $69.35^{+0.61}_{-0.63}$ & $0.299^{+0.017}_{-0.021}$ & $0.167^{+0.080}_{-0.154}$ & -- & $-19.40^{+0.21}_{-0.16}$ \\ 
		CC + $\mathrm{PN}^+$\,\&\,SH0ES + BAO + RSD & $69.38^{+0.67}_{-0.68}$ & $0.282\pm 0.011$ & $0.275^{+0.083}_{-0.096}$ & $0.793\pm 0.035$ & $-19.37^{+0.31}_{-0.30}$ \\ 
		\hline
    \end{tabular}
    }
       \caption{Exact results for $f_2$ model that include the parameters $H_0$, $\Omega_{m,0}$ and $\frac{1}{p_2}$. The $\sigma_{8,0}$ parameter and the nuisance parameter $M$, are provided for data sets that include RSD or $\mathrm{PN}^+$\,\&\,SH0ES, respectively  otherwise, they are left empty.}
    \label{tab:LM}
\end{table}
\begin{figure}[h!]
    \centering
    \includegraphics[width = 0.8\textwidth]{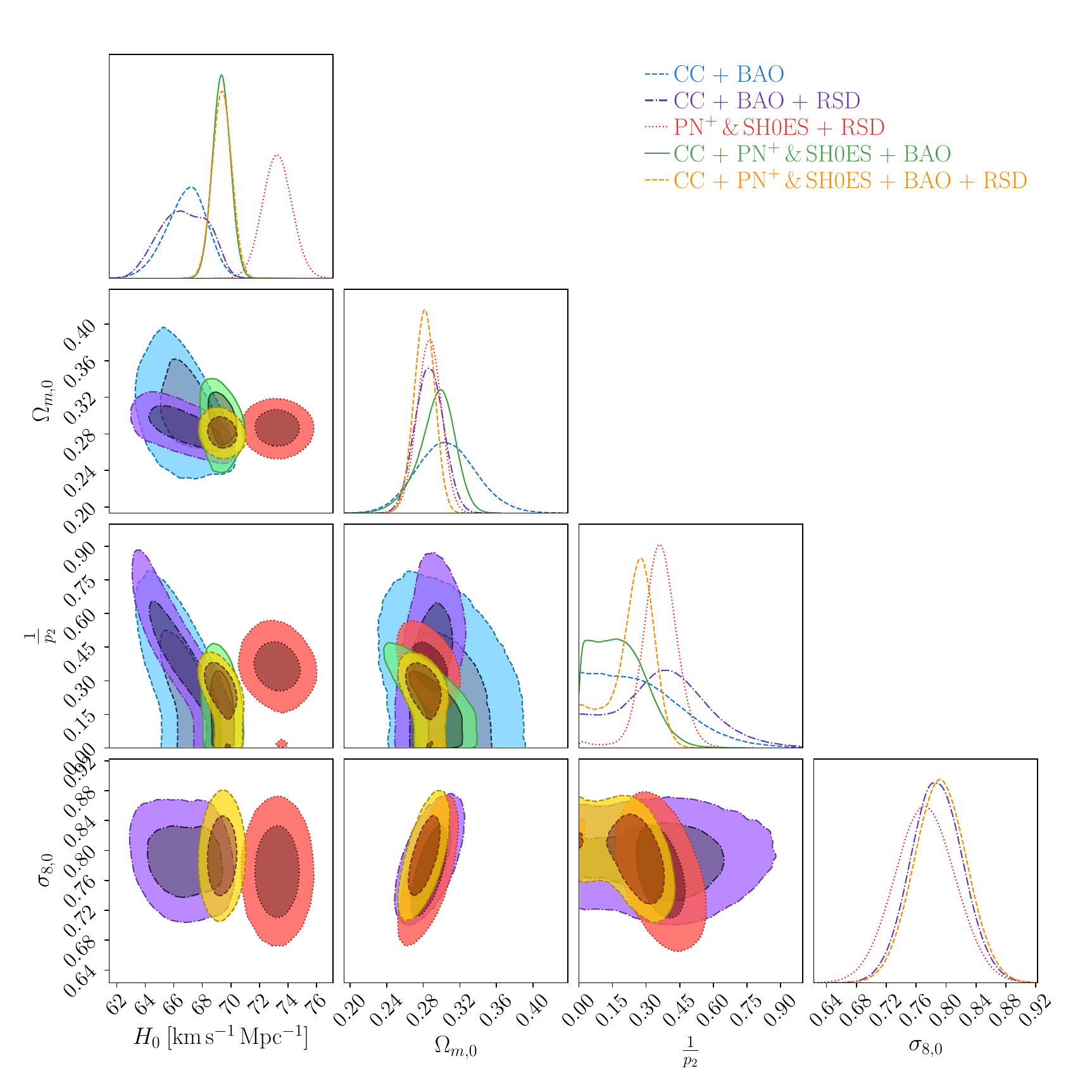}
    \caption{Confidence contours and posterior distributions for the $f_2$CDM model (Linder Model) parameters, including $H_0$, $\Omega_{m,0}$, and $\frac{1}{p_2}$. In cases where the RSD data is incorporated (purple, red, and yellow contours), the $\sigma_{8,0}$ parameter is also displayed.}
    \label{fig:LM}
\end{figure}
Contrary to the $f_1$CDM model, this model can be reduced to the $\Lambda$CDM when $p_2 \rightarrow \infty$. However, for numerical stability, the analysis is conducted with the reciprocal of $p_2$, in such a way that the limit effectively becomes $1/p_2 \rightarrow 0^+$. 

The posterior distributions and confidence levels of the constrained parameters are depicted in Fig.~\ref{fig:LM}. Similar to the $f_1$CDM model, the highest value for the $H_0$ parameter is obtained when combining $\mathrm{PN}^++$RSD data as indicated by the precise values in Table~\ref{tab:LM}. However, in this case, the lowest $H_0$ value, specifically $H_0=66.5^{+2.2}_{-1.3} {\rm\, km \, s}^{-1} {\rm Mpc}^{-1}$, is obtained for the CC+BAO+RSD combination. Overall, in this scenario, the parameter values trend slightly lower compared to the $f_1$CDM model.

Regarding the $\Omega_{m,0}$ parameter, we observe a similar trend to that of the $H_0$ parameter, with lower values being reported. However, the consistent pattern persists, where data sets that include RSD values yield lower values for $\Omega_{m,0}$ compared to their counterparts that do not incorporate RSD data. The anti-correlation between the $H_0$ parameter and the $\Omega_{m,0}$ parameter is still visible.  However, in comparison to the $f_1$CDM model, the anti-correlation between $\Omega_{m,0}$ and the model parameter is not as pronounced, resulting in a higher degree of degeneracy between these two parameters.

The $1/p_2$ parameter values obtained are slightly higher than those in the $f_1$CDM model. Additionally, in contrast to the previous model, the $\Lambda$CDM limit does not fall within the 1$\sigma$ region of the $\Lambda$CDM, indicating a slight deviation from the $\Lambda$CDM model.

Furthermore, the parameter $\sigma_{8,0}$ continues to exhibit the same trend observed previously, with lower values consistently reported. Notably, we observe a correlation between this parameter and $\Omega_{m,0}$ across all data sets. However, the degeneracy between $\sigma_{8,0}$ and $H_0$ remains valid in this model. Similar trend to previous model is also seen with regards $G_{\mathrm{eff}}$, where $G_{\mathrm{eff}} \approx G_N$ where
\begin{equation}
    G_{\mathrm{eff}} = \frac{G_N}{1+\frac{1}{2}\alpha_2 p_2 \sqrt{\frac{T_0}{T}} \mathrm{Exp}\left[-p_2 \sqrt{\frac{T}{T_0}}\right]} \,.
\end{equation}

\begin{figure}[]
  \begin{minipage}{0.5\textwidth}
  \captionsetup{width=0.9\textwidth}
    \includegraphics[width = 0.8\textwidth]{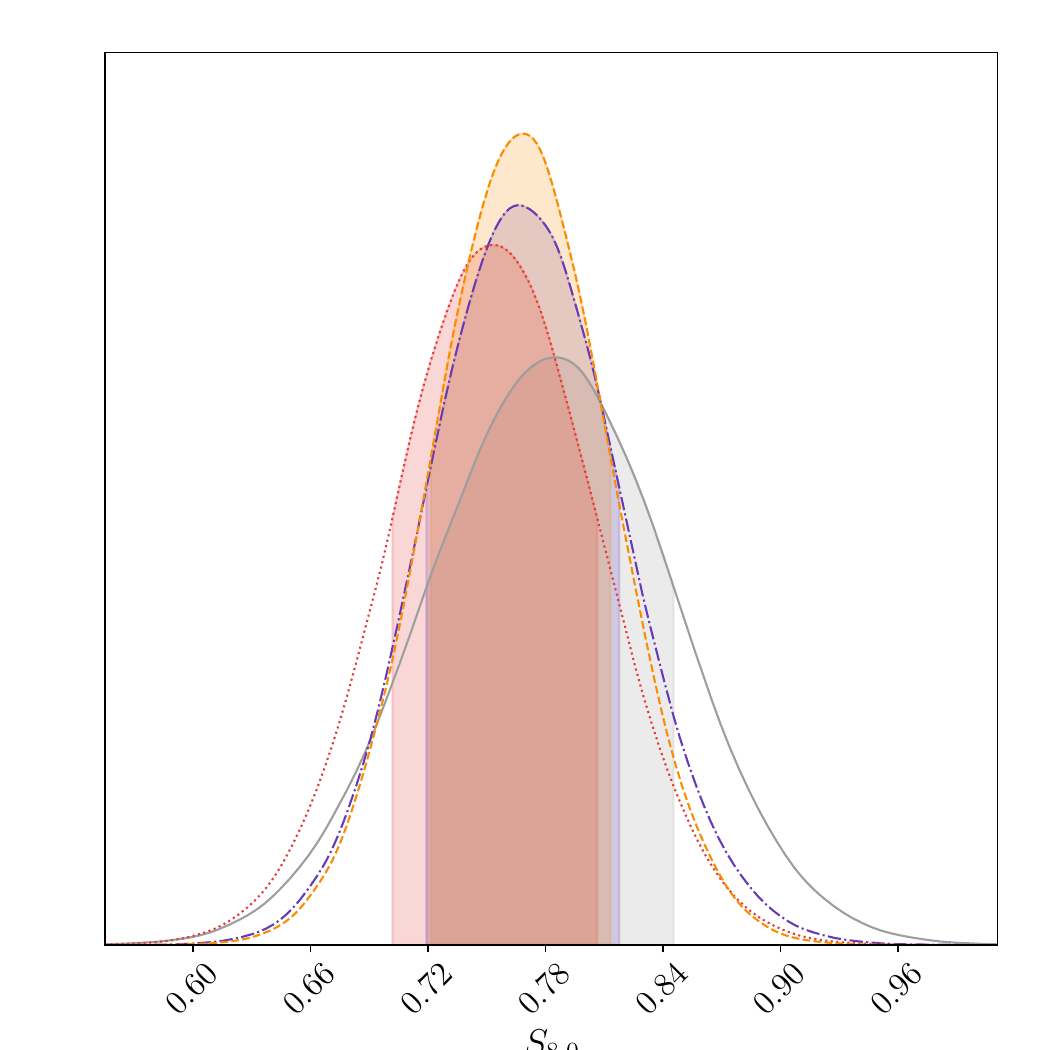}
    \captionof{figure}{Posterior distribution for the $S_{8,0}$ parameter in the $f_2$CDM model. Legend: Grey denotes the RSD data, purple corresponds to CC+BAO+RSD data, red represents the $\mathrm{PN}^+$\,\&,SH0ES + RSD dataset, while orange indicates CC + $\mathrm{PN}^+$\,\&\,SH0ES + BAO + RSD data.}
    \label{fig:LM_S80}
  \end{minipage}
  \begin{minipage}{0.5\textwidth}
  \captionsetup{width=0.9\textwidth}
    \begin{tabular}{lc}
		\hline
  	Data sets & $S_{8,0}$ \\ 
		\hline
		 \textcolor{mygray}{\tikz[baseline=-0.75ex]\draw [thick,solid] (0,0) -- (0.5,0); RSD} & $0.784^{+0.061}_{-0.065}$ \\ 
		\texttransparent{0.7}{\textcolor{patriarch}{\tikz[baseline=-0.75ex]\draw [thick,dash dot dot] (0,0) -- (0.5,0);CC +  BAO + RSD}} & $0.765^{+0.052}_{-0.046}$ \\ 
		\texttransparent{0.7}{\textcolor{myred}{\tikz[baseline=-0.75ex]\draw [thick,dotted] (0,0) -- (0.5,0);$\mathrm{PN}^+$\,\&\,SH0ES + RSD}} & $0.753^{+0.054}_{-0.051}$ \\ 
		\texttransparent{0.8}{\textcolor{myorange}{\tikz[baseline=-0.75ex]\draw [thick,dashed] (0,0) -- (0.5,0);CC + $\mathrm{PN}^+$\,\&\,SH0ES + BAO + RSD}} & $0.768^{+0.045}_{-0.046}$ \\ 
		\hline
    \end{tabular}
    \captionof{table}{Exact $S_{8,0}$ values corresponding to various data sets for the $f_2$CDM model.}
    \label{tab:LM_S80}
  \end{minipage}
\end{figure}

In the cases where RSD is included, we once again calculate the quantity $S_{8,0}$, and we obtain slightly different results compared to the previous model as shown in the Table~\ref{tab:LM_S80} and Fig.~\ref{fig:LM_S80}. Similar results are obtained for the CC+BAO+RSD data sets, but we observe a lower value for the $\mathrm{PN}^+$\,\&\,SH0ES+RSD combination. Conversely, we obtain higher values for the RSD data set on its own and for CC + $\mathrm{PN}^+$\,\&\,SH0ES + BAO + RSD, with the maximum value being achieved for the former data set, whereas previously it had exhibited the minimum value. Lastly, we again check the correlation between the parameters $p$ and $S_{8,0}$ where the degeneracy in the RSD data persists, however, when the $\mathrm{PN}^+$\,\&\,SH0ES data is combined with the RSD data, we now notice an anti-correlation that was not previously evident. A visual representation of these findings can be found in Appendix~\ref{sec:App_pvsS}, specifically in Fig.~\ref{fig:PLM_p_vs_S80}.

\subsection{Exponential Model}
\begin{table}
\resizebox{\textwidth}{!}{%
\centering  
 %   \hspace{-2cm}
    \begin{tabular}{cccccc}
        \hline
		Data Sets & $H_0 \mathrm{\hspace{0.15cm}[km \hspace{0.1cm} s^{-1} \hspace{0.1cm}Mpc ^{-1}]}$ & $\Omega_{m,0}$ & $\frac{1}{p_3}$ & $\sigma_{8,0}$ & $M$ \\ 
		\hline
		CC + BAO  & $67.5^{+1.7}_{-2.3}$ & $0.311^{+0.039}_{-0.034}$ & $0.058^{+0.182}_{-0.056}$ & -- & -- \\ 
		CC +  BAO + RSD & $68.6^{+1.3}_{-1.9}$ & $0.276^{+0.016}_{-0.015}$ & $0.026^{+0.214}_{-0.025}$ & $0.798^{+0.040}_{-0.036}$ & -- \\ 
		$\mathrm{PN}^+\,\&\,$SH0ES + RSD & $73.2^{+1.0}_{-1.1}$ & $0.280^{+0.014}_{-0.015}$ & $0.232^{+0.027}_{-0.031}$ & $0.793^{+0.038}_{-0.039}$ & $-19.25\pm 0.11$ \\ 
		CC + $\mathrm{PN}^+\,\&\,$SH0ES + BAO  & $69.34^{+0.65}_{-0.64}$ & $0.300^{+0.017}_{-0.016}$ & $0.160^{+0.029}_{-0.126}$ & -- & $-19.34^{+0.24}_{-0.31}$ \\ 
		CC + $\mathrm{PN}^+\,\&\,$SH0ES + BAO + RSD & $69.54^{+0.64}_{-0.66}$ & $0.282\pm 0.012$ & $0.197^{+0.038}_{-0.092}$ & $0.807\pm 0.032$ & $-19.38^{+0.20}_{-0.19}$ \\ 
		\hline
    \end{tabular}
    }
    \caption{Exact results for $f_3$ model that include the parameters $H_0$, $\Omega_{m,0}$ and $\frac{1}{p_3}$. The $\sigma_{8,0}$ parameter and the nuisance parameter $M$, are provided for data sets that include RSD or $\mathrm{PN}^+$\,\&\,SH0ES, respectively  otherwise, they are left empty.}
    \label{tab:LM2}
\end{table}
 
The next model under consideration in this analysis is the exponential model, hereafter $f_3$CDM, which draws inspiration from previous works on $f(\lc{R})$ \cite{Linder:2009jz}. In fact, a variant of the Linder model is proposed in Ref.~\cite{Nesseris:2013jea}, where the square root in the exponential form is no longer present. In this case, $\mathcal{F}_3$ is expressed as an exponential function with two model constants, $\alpha_3$ and $p_3$, along with the current torsion scalar $T_0$ and the variable $T$, such that
\begin{equation}
    \mathcal{F}_3 = \alpha_3 T_0 \left(1- \mathrm{Exp}\left[-p_3 T/T_0\right] \right) \,.
\end{equation}

By evaluating the Friedmann equation at the present time, we can determine the value of the constant $\alpha_3$, which is calculated as
\begin{equation} \label{eq:alpha3}
    \alpha_3 = \frac{1-\Omega_{m,0} - \Omega_{r,0}}{(1+2p_3)e^{-p_3}-1}\,.
\end{equation}

Therefore, by substituting Eq.~\eqref{eq:alpha3} into the modified Friedmann Equation, Eq.~\eqref{eq:Friedmann_1}, we can derive the Friedmann equation for this model, which can be solved numerically
\begin{equation}
    E^2\left(z\right) = \Omega_{m,0} \left(1+z\right)^3 + \Omega_{r_0}\left(1+z\right)^4 + \frac{1 - \Omega_{m,0} - \Omega_{r_0}}{(1 + 2p_3 )e^{-p_3} - 1} \left[\left(1 + 2p_3 E^2 (z)\right)\text{Exp}\left[-p_3 E^2 (z)\right] - 1\right]\,.
\end{equation}

This model exhibits behavior similar to the Linder Model, where the $\Lambda$CDM limit is approached as $p_3 \rightarrow \infty$. Therefore, as previously discussed, we perform the analyses using $1/p_3$ to ensure numerical stability, as this approach aligns with the previous model. In this case the $G_{\mathrm{eff}}$ is defined as 
\begin{equation}
    G_{\mathrm{eff}} = \frac{G_N}{1 + \alpha_3 p_3 \mathrm{Exp}\left[-p_3 \frac{T}{T_0}\right]} \,,
\end{equation}
where this model exhibits a similar trend to the previous ones.

\begin{figure}[]
    \centering
    \includegraphics[width = 0.8\textwidth]{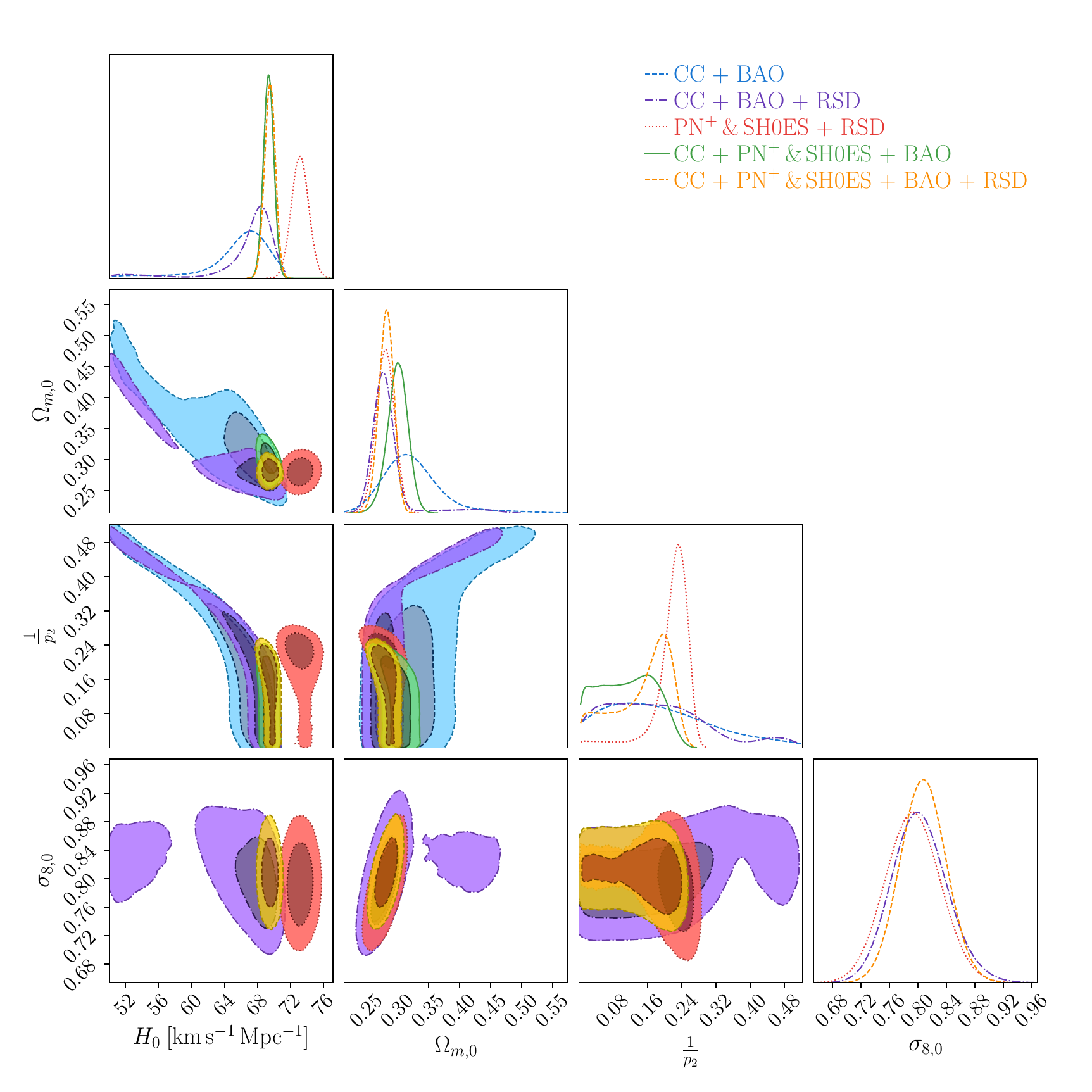}
    \caption{Confidence contours and posterior distributions for the $f_3$CDM model (Exponential Model) parameters, including $H_0$, $\Omega_{m,0}$, and $\frac{1}{p_3}$. In cases where the RSD data is incorporated (purple, red, and yellow contours), the $\sigma_{8,0}$ parameter is also displayed.}
    \label{fig:LM2}
\end{figure}

The confidence levels and the posteriors are found in Fig.~\ref{fig:LM2}, whilst the constrained values are found in Table~\ref{tab:LM2}. Removing the square root component has notably influenced the constraints, particularly on $\sigma_{8,0}$. The $\sigma_{8,0}$ values are considerably higher in this model compared to those reported in $f_2$CDM.

As with $f_1$CDM, the model $f_3$CDM exhibits the highest constrained value for $H_0$ when considering the $\mathrm{PN}^+$\,\&\,SH0ES+RSD data ($H_0 = 73.2^{+1.1}_{-1.14} {\rm\, km \, s}^{-1} {\rm Mpc}^{-1}$) combination, while the lowest value is obtained for CC+BAO data ($H_0 = 67.5^{+1.7}_{-2.3} {\rm\, km \, s}^{-1} {\rm Mpc}^{-1}$). Similarly, the density parameter $\Omega_{m,0}$ also shows the highest value for CC+BAO ($\Omega_{m,0} = 0.311^{+0.039}_{-0.034}$) and the lowest for CC+$\mathrm{PN}^+$\,\&\,SH0ES+BAO ($\Omega_{m,0} = 0.211 \pm 0.012$). In addition, the RSD data appears to impose more stringent constraints on these parameters, particularly on the density parameter.

Regarding the model parameter, unlike in the case of $f_2$CDM, we observe that its range within the $1 \sigma$ and $2 \sigma$ confidence intervals is narrower. In this instance, the $2 \sigma$ interval spans from 0 to a maximum of 0.5, but akin in $f_2$CDM still lie within 2$\sigma$ of the $\Lambda$CDM limit. 

\begin{figure}[]
  \begin{minipage}{0.5\textwidth}
  \captionsetup{width=0.9\textwidth}
    \includegraphics[width = 0.8\textwidth]{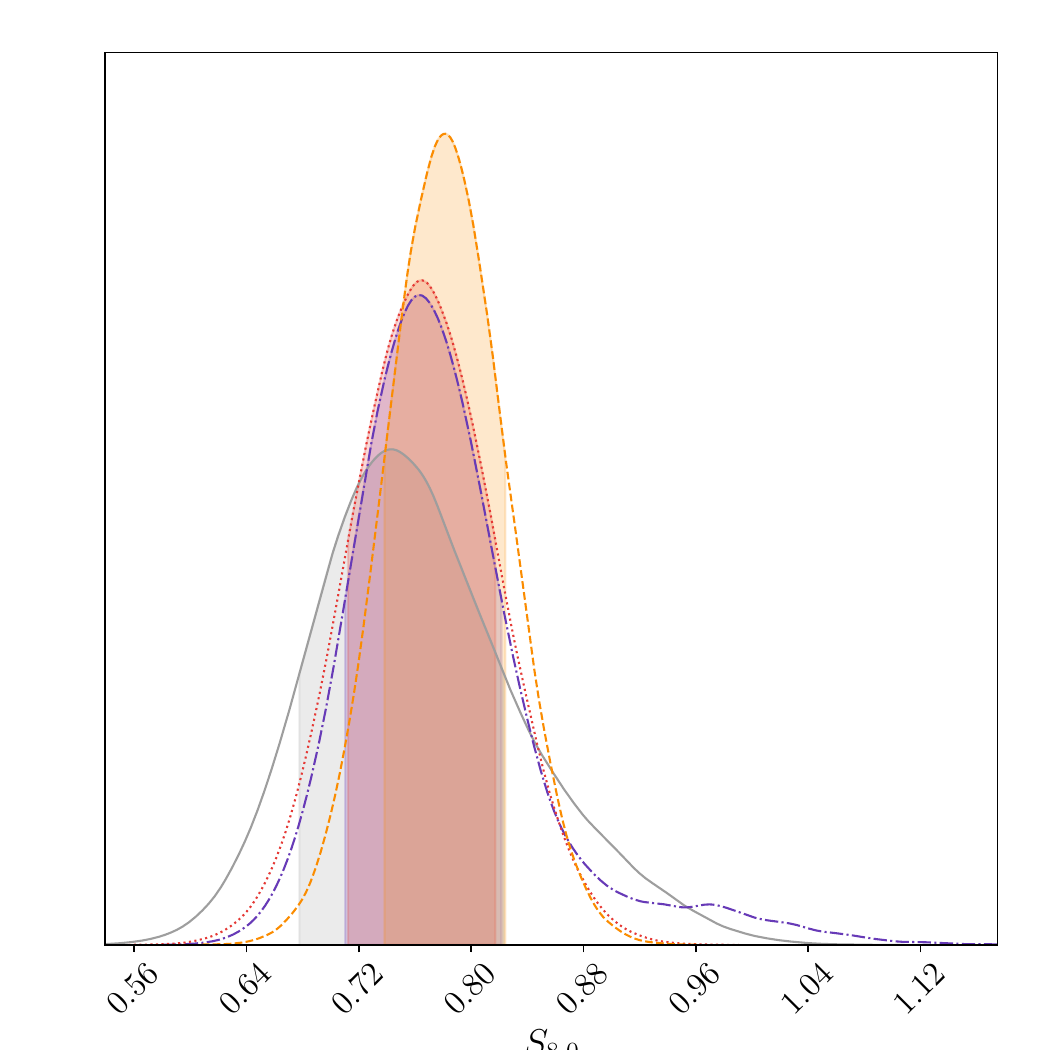}
    \captionof{figure}{Posterior distribution for the $S_{8,0}$ parameter in the $f_3$CDM model. Legend: Grey denotes the RSD data, purple corresponds to CC+BAO+RSD data, red represents the $\mathrm{PN}^+$\,\&,SH0ES + RSD dataset, while orange indicates CC + $\mathrm{PN}^+$\,\&\,SH0ES + BAO + RSD data.}
    \label{fig:LM2_S80}
  \end{minipage}
  \begin{minipage}{0.5\textwidth}
  \captionsetup{width=0.9\textwidth}
    \begin{tabular}{lc}
        \hline
		Data Sets & $S_{8,0}$ \\ 
		\hline

		\textcolor{mygray}{\tikz[baseline=-0.75ex]\draw [thick,solid] (0,0) -- (0.5,0); RSD} & $0.744^{+0.079}_{-0.066}$ \\ 
		\texttransparent{0.7}{\textcolor{patriarch}{\tikz[baseline=-0.75ex]\draw [thick,dash dot dot] (0,0) -- (0.5,0);CC +  BAO + RSD}} & $0.762^{+0.059}_{-0.052}$ \\ 
		\texttransparent{0.7}{\textcolor{myred}{\tikz[baseline=-0.75ex]\draw [thick,dotted] (0,0) -- (0.5,0);$\mathrm{PN}^+$\,\&\,SH0ES + RSD}} & $0.765^{+0.053}_{-0.052}$ \\ 
		\texttransparent{0.8}{\textcolor{myorange}{\tikz[baseline=-0.75ex]\draw [thick,dashed] (0,0) -- (0.5,0);CC + $\mathrm{PN}^+$\,\&\,SH0ES + BAO + RSD}} & $0.782\pm 0.043$ \\ 
		\hline
    \end{tabular}
    \captionof{table}{Exact $S_{8,0}$ values corresponding to various data sets for the $f_3$CDM model.}
    \label{tab:LM2_S80}
  \end{minipage}
\end{figure}

However, the most significant difference in this model becomes evident in the $\sigma_{8,0}$ parameter. This model reports a higher value for this parameter. Furthermore, a correlation between the parameters $\Omega_{8,0}$ and $\sigma_{8,0}$ is now evident, which was not observed in the $f_1$CDM model but is apparent in the $f_2$CDM model. 
The degeneracy between the $H_0$ and $\sigma_{8,0}$ parameters, however, remains apparent, as observed in the other cases.
The highest value for $\sigma_{8,0}$ is obtained for the CC + $\mathrm{PN}^+$\,\&\,SH0ES + BAO + RSD combination, yielding a value of $\sigma_{8,0} = 0.793 \pm 0.035$, as detailed in Table~\ref{tab:LM2_S80} and illustrated in Fig.~\ref{fig:LM2_S80}. Conversely, the lowest value for $S_{8,0}$ is achieved when using the $\mathrm{PN}^+$\,\&\,SH0ES+RSD data, as indicated in both tables presenting the $\sigma_{8,0}$ and $S_{8,0}$ values respectively. Additionally, consistent with previous models, the value for $S_{8,0}$ obtained using the RSD data alone is relatively low, akin to the $f_1$CDM model.

The final piece of analysis for the $f_3$CDM model is depicted in Fig.~\ref{fig:PLM_p_vs_S80} in Appendix~\ref{sec:App_pvsS}. Here, we observe a degenerate relationship between the parameters $p_3$ and $S_{8,0}$, which transitions into an anti-correlation at higher values of $p_3$. This effect is particularly noticeable for the data sets that incorporate BAO measurements.

\subsection{Logarithmic Model}

The last model, hereafter referred to as $f_4$CDM, which was proposed by Bamba et al. \cite{Bamba:2010wb}, is known as the logarithmic model. This model is characterized by the following expression
\begin{equation}\label{eq:LogM}
    \mathcal{F}_4 (T) = \alpha_4 T_0 \sqrt{\frac{T}{p_4 T_0}} \log\left[\frac{p_4 T_0}{T}\right] \,,
\end{equation}
where $\alpha_4$ and $p_4$ are the two model constants. To determine the constant $\alpha_4$, we follow the standard procedure of evaluating the Friedmann equation, Eq.~\eqref{eq:Friedmann_2}, at the current time, resulting in
\begin{equation}
    \alpha_4 = -\frac{\left(1-\Omega_{m,0} - \Omega_{r,0} \right)\sqrt{p_4}}{2} \,.
\end{equation}
This simplifies the Friedmann equation to a more straightforward form
\begin{equation} \label{eq:LogM_FE}
    E^2\left(z\right) = \Omega_{m_0} \left(1+z\right)^3 + \Omega_{r_0}\left(1+z\right)^4 + \left(1 - \Omega_{m,0} - \Omega_{r_0} \right) E(z) \,.
\end{equation}

\begin{figure}[]
    \centering
    \includegraphics[width = 0.8\textwidth]{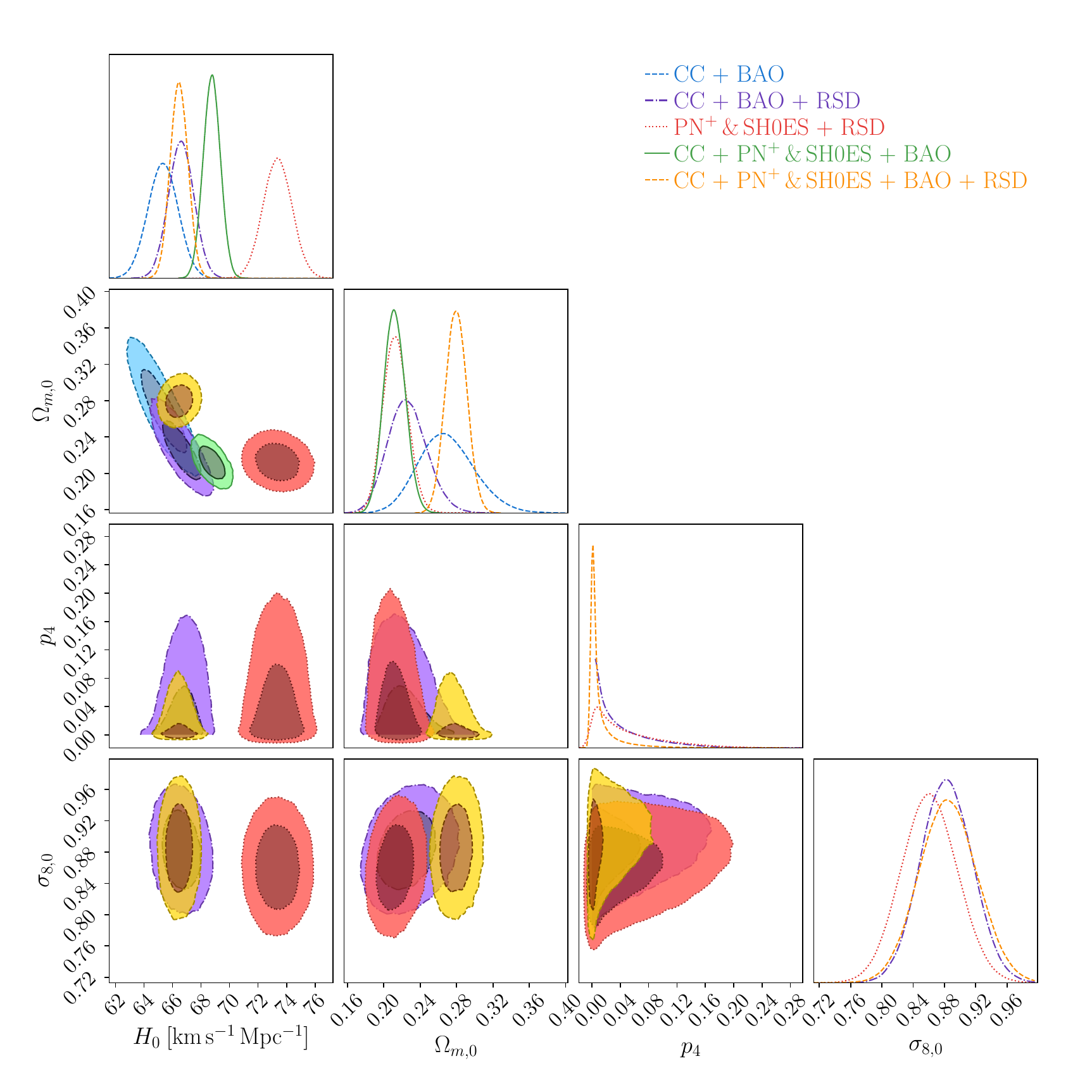}
    \caption{Confidence contours and posterior distributions for the $f_4$CDM model (Logarithmic Model) parameters, including $H_0$ and $\Omega_{m,0}$. In cases where the RSD data is incorporated (purple, red, and yellow contours), the $\sigma_{8,0}$ and $p_4$ parameter is also displayed.}
    \label{fig:LogM}
\end{figure}

Interestingly, the additional model parameter $p_4$ does not feature in Eq.~\eqref{eq:LogM_FE}, indicating that at background level this parameter cannot be constrained. Nonetheless, it is noteworthy that this parameter does come into play at the perturbative level, particularly in the context of linear matter perturbations, as illustrated in Eq.~\eqref{eq:delta_prime2}, where $p_4$ is featured in the $G_{\mathrm{eff}}$ term, in particular in the $\mathcal{F}_T$ term. Therefore, as can be seen in the contour plots in Fig,~\ref{fig:LogM} and Table|~\ref{tab:LogM}, the $p_4$ parameter does not feature in the CC+BAO (blue contours) and CC+$\mathrm{PN}^+$\,\&\,SH0ES+BAO (green contour),  but is relevant for the rest of the data sets that include the RSD data.

The inclusion of the logarithmic function makes a noticeable difference in this model, particularly when examining the relationship between the $H_0$ and $\Omega_{m,0}$ parameters. In this case, the relationship exhibits a distinct anti-correlation, which is more pronounced in this model compared to the previous ones. Interestingly, this results in more extreme values for the $H_0$ parameter, with the highest value achieved being $H_0 = 73.40^{+0.97}_{-1.06} {\rm\, km \, s}^{-1} {\rm Mpc}^{-1}$ for the $\mathrm{PN}^+$\,\&\,SH0ES data set. Conversely, the lowest value was obtained for the CC+BAO data set, with a value of $H_0 = 65.3^{+1.1}_{-1.0} {\rm\, km \, s}^{-1} {\rm Mpc}^{-1}$

\begin{table}
\resizebox{\textwidth}{!}{%
\centering  
 %   \hspace{-2cm}
    \begin{tabular}{cccccc}

        \hline
		Data Sets & $H_0 \mathrm{\hspace{0.15cm}[km \hspace{0.1cm} s^{-1} \hspace{0.1cm}Mpc ^{-1}]}$ & $\Omega_{m,0}$ & $p_4$ & $\sigma_{8,0}$ & $M$ \\ 
		\hline
		CC + BAO  & $65.3^{+1.1}_{-1.0}$ & $0.266^{+0.031}_{-0.030}$ & -- & -- & -- \\ 
		CC +  BAO + RSD & $66.62^{+0.87}_{-0.91}$ & $0.224^{+0.022}_{-0.020}$ & $0.005^{+0.031}_{-0.000}$ & $0.883\pm 0.033$ & -- \\ 
		$\mathrm{PN}^+\,\&\,$SH0ES + RSD & $73.40^{+0.97}_{-1.06}$ & $0.213^{+0.013}_{-0.014}$ & $0.007^{+0.057}_{-0.013}$ & $0.860^{+0.036}_{-0.035}$ & $-19.09^{+0.16}_{-0.23}$ \\ 
		CC + $\mathrm{PN}^+\,\&\,$SH0ES + BAO  & $68.78^{+0.58}_{-0.63}$ & $0.211\pm 0.012$ & -- & -- & $-19.388\pm 0.017$ \\ 
		CC + $\mathrm{PN}^+\,\&\,$SH0ES + BAO + RSD & $66.41^{+0.67}_{-0.56}$ & $0.280^{+0.011}_{-0.012}$ & $0.0015^{+0.009}_{-0.005}$ & $0.883^{+0.038}_{-0.036}$ & $-19.57^{+0.16}_{-0.15}$ \\ 
		\hline
    \end{tabular}
    }
    \caption{Exact results for $f_4$ model that include the parameters $H_0$ and $\Omega_{m,0}$. The $\sigma_{8,0}$ and $p_4$ parameter together with the nuisance parameter $M$, are provided for data sets that include RSD or $\mathrm{PN}^+$\,\&\,SH0ES, respectively  otherwise, they are left empty.}
    \label{tab:LogM}
     \hspace{-2cm}
\end{table}

The values of $\Omega_{m,0}$ also exhibit significant variations. In this case, the lowest value is achieved for the CC+$\mathrm{PN}^+$\,\&\,SH0ES+BAO data set, while the highest is for the same data set but with the inclusion of the RSD data (i.e CC+$\mathrm{PN}^+$\,\&\,SH0ES+BAO+RSD). It seems like when the RSD model is included in the data sets, either the $H_0$ parameter or the $\Omega_{m,0}$ is driven to higher values when compared to its counterpart data set without the RSD data set. Therefore, in this model, it appears that the presence of RSD data influences the understanding of the Universe's current acceleration and the role of baryonic and dark matter components within this model. This influence can be attributed to RSD data's exceptional sensitivity to the distribution of matter, which in turn plays a crucial role in shaping the large-scale structure of the Universe through its gravitational interactions.

A particularly intriguing aspect of this model, which sets it apart from the $f_1-f_3$CDM models, is the absence of a $\Lambda$CDM limit both at the background and perturbative levels. At the perturbative level, the parameter $p_4$ appears in $G_\mathrm{eff}$, which is expressed as
\begin{equation}
    G_\mathrm{eff} = \frac{G_N}{1 + \frac{\alpha_4}{2\sqrt{p_4}} \left(\frac{T}{T_0}\right)^{-\frac{1}{2}} \ln\left(\frac{p_4T_0}{T}\right) + \alpha_4 \sqrt{\frac{T_0}{p_4T}}} \,.
\end{equation}
In this scenario, no choice of the value of $p_4$ can reproduce a $\Lambda$CDM model. Furthermore, given that it appears in the logarithmic function, the parameter $p_4$ must be greater than zero, as illustrated in Fig.~\ref{fig:LogM}. However, Table~\ref{tab:LogM}, still shows that the value of $p_4$ is very close to zero, which makes $G_\mathrm{eff}$, slightly higher than $G_N$, throughout the redshift span. 

\begin{figure}[]
  \begin{minipage}{0.45\textwidth}
  \captionsetup{width=0.9\textwidth}
  \centering
    \includegraphics[width = 0.8\textwidth]{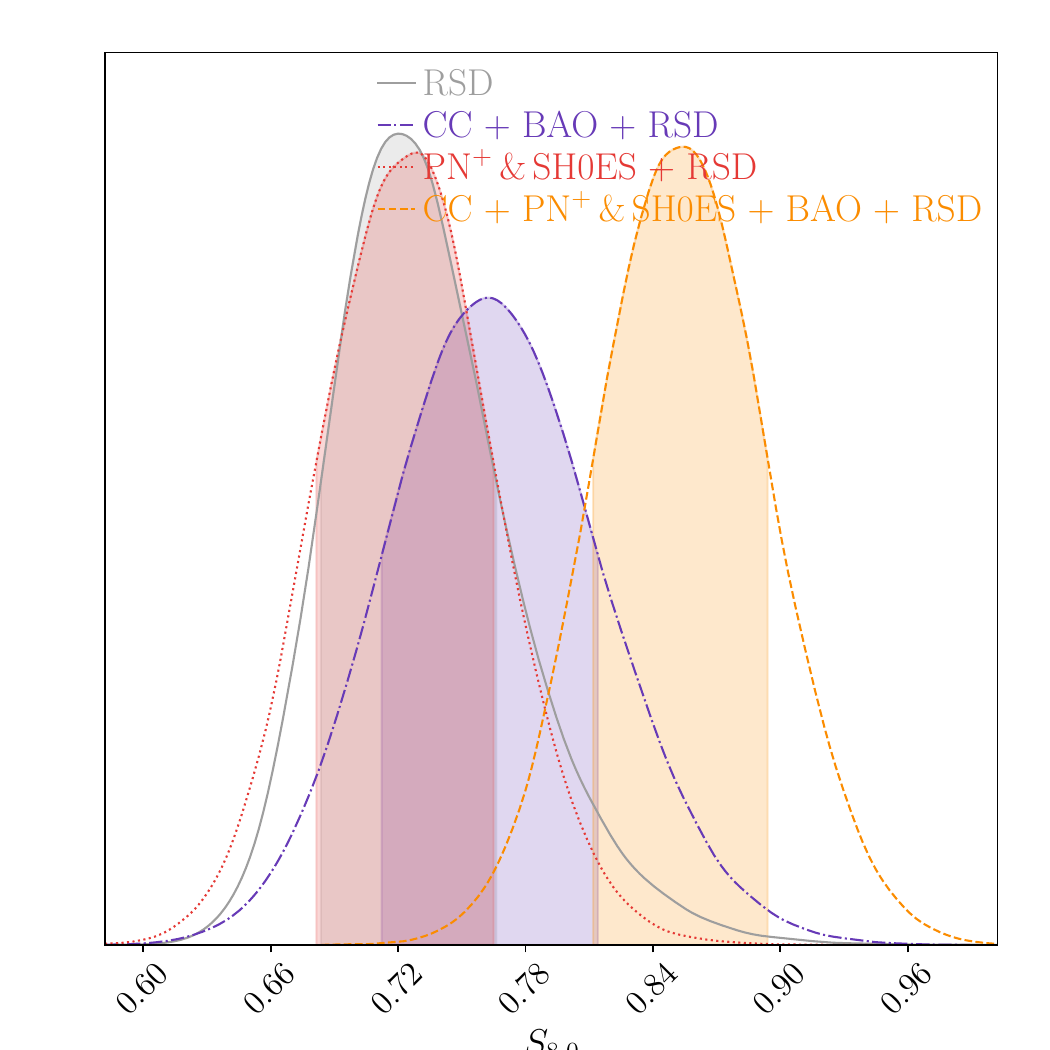}
    \captionof{figure}{Posterior distributionfor the $S_{8,0}$ parameter in the $f_4$CDM model. Legend: Grey denotes the RSD data, purple corresponds to CC+BAO+RSD data, red represents the $\mathrm{PN}^+$\,\&,SH0ES + RSD dataset, while orange indicates CC + $\mathrm{PN}^+$\,\&\,SH0ES + BAO + RSD data.}
    \label{fig:LogM_S80}
  \end{minipage}
   \begin{minipage}{0.5\textwidth}
   \captionsetup{width=0.9\textwidth}
    \begin{tabular}{lc}
        \hline
		Data Sets  & $S_{8,0}$ \\ 
		\hline
		\textcolor{mygray}{\tikz[baseline=-0.75ex]\draw [thick,solid] (0,0) -- (0.5,0); RSD} & $0.720^{+0.047}_{-0.036}$ \\ 
		\texttransparent{0.7}{\textcolor{patriarch}{\tikz[baseline=-0.75ex]\draw [thick,dash dot dot] (0,0) -- (0.5,0);CC +  BAO + RSD}} & $0.762^{+0.052}_{-0.050}$ \\ 
		\texttransparent{0.7}{\textcolor{myred}{\tikz[baseline=-0.75ex]\draw [thick,dotted] (0,0) -- (0.5,0);$\mathrm{PN}^+$\,\&\,SH0ES + RSD}} & $0.730^{+0.035}_{-0.048}$ \\ 
		\texttransparent{0.8}{\textcolor{myorange}{\tikz[baseline=-0.75ex]\draw [thick,dashed] (0,0) -- (0.5,0);CC + $\mathrm{PN}^+$\,\&\,SH0ES + BAO + RSD}} & $0.854^{+0.040}_{-0.042}$ \\ 
		\hline
    \end{tabular}
    \captionof{table}{Exact $S_{8,0}$ values corresponding to various data sets for the $f_4$CDM model.}
    \label{tab:LogM_S80}
  \end{minipage}
\end{figure}

The $\sigma_{8,0}$ parameter in this model appears to be significantly higher when compared to the other models. Interestingly, the CC+BAO+RSD and CC+$\mathrm{PN}^+$\,\&\,SH0ES+BAO+RSD data sets both report the same value for $\sigma_{8,0}$, suggesting that the inclusion of $\mathrm{PN}^+$\,\&\,SH0ES data has a minimal effect on this parameter in this particular model. The CC+$\mathrm{PN}^+$\,\&\,SH0ES + RSD data set reports a slightly lower value for $\sigma_{8,0}$. This observation leads us to consider Fig.~\ref{fig:LogM_S80} and Table~\ref{tab:LogM_S80}, where we can see the values obtained for this parameter. In this case, it's evident that $\Omega_{m,0}$ has a notable impact on this parameter. While $\sigma_{8,0}$ is the same for CC+BAO+RSD and CC+$\mathrm{PN}^+$\,\&\,SH0ES+BAO+RSD, $S_{8,0}$ varies significantly between the two data sets. A higher value is obtained when $\mathrm{PN}^+$\,\&\,SH0ES data is included, making it the data set with the highest $S_{8,0}$ value.

The last Fig.~\ref{fig:PLM_p_vs_S80} (Appendix~\ref{sec:App_pvsS}), shows the relationship between the $S_{8,0}$ and $p_4$. As expected based on previous models, we observe a degeneracy between these parameters, particularly concentrated within the 1$\sigma$ region near $p_4=0$.

\section{Analysis}\label{sec:analysis}

To evaluate how well each $f_i$CDM model performs with various data sets, we employ several statistical measures. Firstly, we calculate the minimum $\chi^2_\mathrm{min}$ values for each model and dataset. These values are derived from the maximum likelihood $L_\mathrm{max}$, with the relationship being $\chi^2_\mathrm{min} = -2 \ln L_\mathrm{max}$. A lower $\chi^2_\mathrm{min}$ indicates a better fit of the model to the data.

In addition to $\chi^2_\mathrm{min}$, we also compare the models to the standard $\Lambda$CDM model using two criteria: the Akaike Information Criterion (AIC) and the Bayesian Information Criterion (BIC). These criteria consider both the model's goodness of fit, represented by $\chi^2_\mathrm{min}$, and its complexity, which is determined by the number of parameters $n$. The AIC is calculated as
\begin{equation}
    \mathrm{AIC} = \chi^2_\mathrm{min} + 2n \,,
\end{equation}
where a lower AIC value indicates a model that fits the data better while accounting for complexity. It penalizes models with more parameters, even if they exhibit a superior data fit. On the other hand, the BIC is given by
\begin{equation}
    \mathrm{BIC} = \chi^2_\mathrm{min} + n \ln m \,,
\end{equation}
where $m$ is the sample size of the observational data combination. Similar to the AIC, the BIC aims to balance data fit against model complexity. However, it imposes a heavier penalty on models with more parameters as the sample size increases. Therefore, by comparing the AIC and BIC values of different models, we can determine which model is better supported by the data. Generally, models with lower AIC and BIC values are preferred, provided the differences are significant.

\begin{table}[]
\centering
\begin{tabular}{c||c|c|c||c|c|c}
 & \multicolumn{3}{c||}{CC+BAO} & \multicolumn{3}{c}{CC+BAO+RSD} \\
 \hline
 \hline
 & $\chi^2_\mathrm{min}$ & $\Delta$AIC & $\Delta$BIC &    $\chi^2_\mathrm{min}$ & $\Delta$AIC & $\Delta$BIC     \\
 \hline
$\Lambda$CDM & 20.93 & 0    &  0   & 37.14   &   0  &  0      \\

$f_1$CDM & 20.87 &  1.94   &  1.61 &  37.04  & 1.91    & 3.15           \\

$f_2$CDM &  20.93   &  2.00   & 1.66    &  35.41  & 0.28    & 1.52           \\

$f_3$CDM &  20.93   &  2.00   &  1.66   &  37.21  & 2.08    &   3.32       \\

$f_4$CDM &  27.95   & 7.02    &   7.02  &  42.03  & 6.89    &  8.14         \\ 

\end{tabular}
\caption{Comparison of $\chi^2_\mathrm{min}$ and differences in AIC and BIC between the models and $\Lambda$CDM (i.e $\Delta$AIC and $\Delta$BIC). On the left-hand side, results are presented for CC+BAO, while the right-hand side includes RSD.}
\label{tab:CB_CBf_AIC_BIC}
\end{table}

\begin{table}[]
\centering
\begin{tabular}{c||c|c|c||c|c|c||c|c|c}
 & \multicolumn{3}{c||}{$\mathrm{PN}^+$\,\&\,SH0ES+RSD} & \multicolumn{3}{c||}{CC+$\mathrm{PN}^+$\,\&\,SH0ES+BAO} & \multicolumn{3}{c}{CC+$\mathrm{PN}^+$\,\&\,SH0ES+BAO+RSD} \\
\hline
\hline
 &   $\chi^2_{\mathrm{min}}$ &   $\Delta$AIC &  $\Delta$BIC & $\chi^2_{\mathrm{min}}$ &   $\Delta$AIC &  $\Delta$BIC & $\chi^2_{\mathrm{min}}$ &   $\Delta$AIC &  $\Delta$BIC     \\
\hline
$\Lambda$CDM & 1550.20 &  0     &  0    &   1572.60    &   0    &   0   &  1590.71     &  0    &  0    \\

$f_1$CDM &   1549.52    &  1.32     &  2.56  &  1572.56     &  1.96     &   1.78   &   1590.56    & 1.85     & 3.10    \\

$f_2$CDM &   1541.46    &  $-6.74$     & $-5.50$     &  1572.50 &  1.89     & 1.72     &   1587.49    & $-1.21$      &  0.04    \\

$f_3$CDM&   1543.13    &   $-5.07$   &  $-3.84$   &  1572.31     &   1.71  &   1.53   &  1588.33     &   $-0.37$    &  0.88    \\

$f_4$CDM &   1539.71    &   $-8.48$    &  $-7.25$    &  1586.67     &  14.07     &   14.07   &   1663.50    &  74.80     & 76.04 \\

\end{tabular}
\caption{Comparison of $\chi^2_\mathrm{min}$ and differences in AIC and BIC between the models and $\Lambda$CDM (i.e $\Delta$AIC and $\Delta$BIC). On the left-hand side, results are presented for $\mathrm{PN}^+$\,\&\,SH0ES+RSD, whilst in the middle CC+$\mathrm{PN}^+$\,\&\,SH0ES+BAO. On the right-hand side the results for CC+$\mathrm{PN}^+$\,\&\,SH0ES+BAO+RSD are displayed.}
\label{tab:Pf_CPB_CPBf_AIC_BIC}
\end{table}

To assess the performance of different models using various combinations of data sets, we calculate the differences in AIC and BIC between each model ($f_i$CDM) and the reference model, which is the $\Lambda$CDM model. This comparison helps us understand how well each model aligns with the standard model of cosmology. The differences in AIC and BIC are denoted as $\Delta$AIC $= \Delta \chi^2_\mathrm{min} + 2 \Delta n$, and $\Delta$BIC $= \Delta \chi^2_\mathrm{min} + \Delta n \ln m$, respectively. These metrics quantify how each model deviates from the reference model (in this case, $\Lambda$CDM) in which, smaller values of $\Delta$AIC and $\Delta$BIC indicate that a model, along with its chosen dataset, is more similar to the $\Lambda$CDM model, suggesting better performance. Table~\ref{tab:CB_CBf_AIC_BIC} provides a comparison of these metrics for two specific data set combinations: CC+BAO and CC+BAO+RSD. Similarly, Table~\ref{tab:Pf_CPB_CPBf_AIC_BIC} allows us to compare the AIC and BIC values for three different sets of data combinations: $\mathrm{PN}^+ \&$SH0ES+RSD, CC+$\mathrm{PN}^+ \&$SH0ES+BAO and CC+$\mathrm{PN}^+ \&$SH0ES+BAO+RSD.

\begin{figure}
    \centering
    \includegraphics[width = 0.7\textwidth]{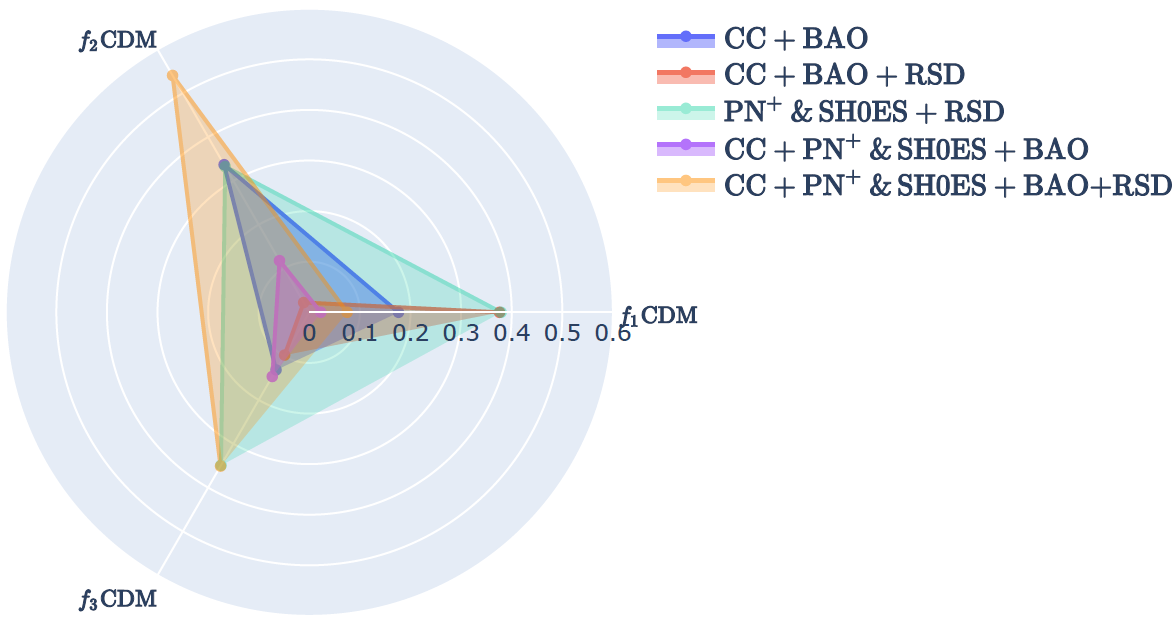}
    \caption{Distances measured in standard deviations ($\sigma$ units) between the constrained $H_0$ values obtained from the $f_{1-3}$CDM models and their corresponding values in the $\Lambda$CDM model. Different colours represent different data sets.}
    \label{fig:Radar_H0_LCDM}
\end{figure}

\begin{figure}[]
    \hspace{-1.2cm}
  \begin{minipage}{0.56\textwidth}
    \includegraphics[width = \textwidth]{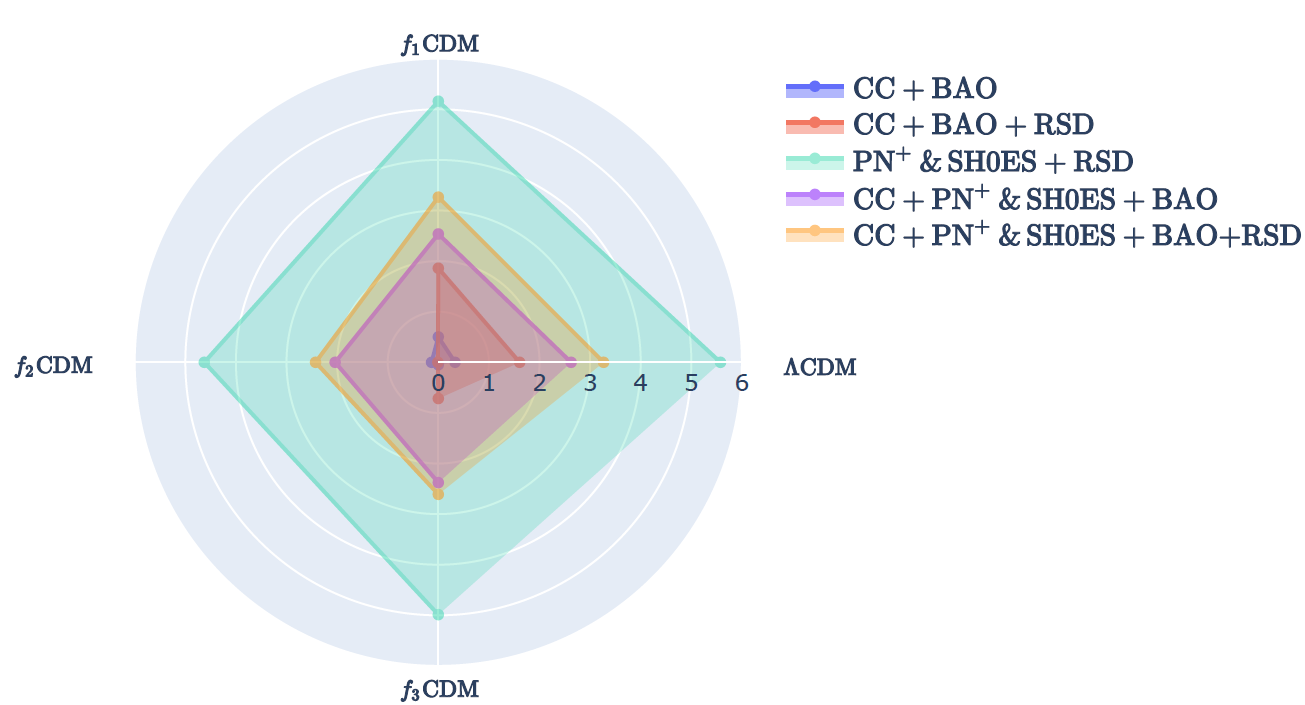}
    % \captionof{figure}{Distances measured in standard deviations ($\sigma$ units) between the constrained $H_0$ values obtained from the $f_{1-3}$CDM models together with $\Lambda$CDM and $H_0^{\mathrm{P18}}$ value. Different colours represent different data sets.}
    \label{fig:Radar_H0_P18}
  \end{minipage}
  \begin{minipage}{0.56\textwidth}
 \includegraphics[width = \textwidth]{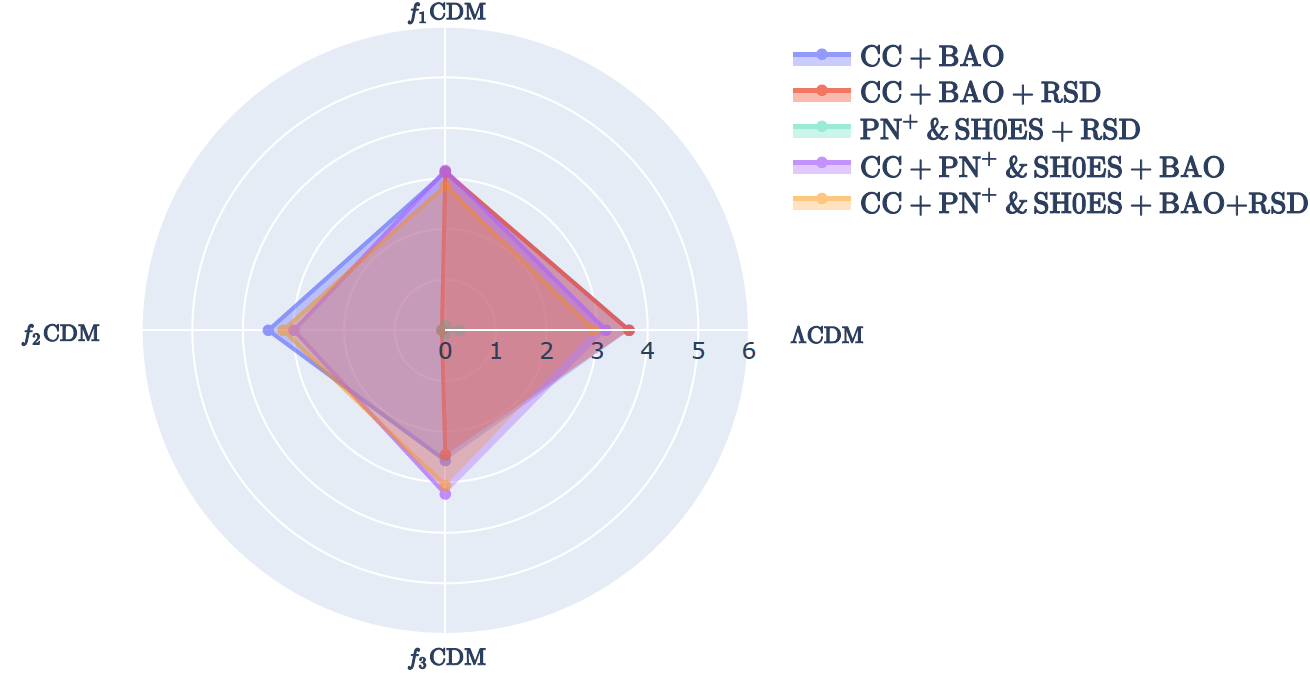}
    % \captionof{figure}{Distances measured in standard deviations ($\sigma$ units) between the constrained $H_0$ values obtained from the $f_{1-3}$CDM models together with $\Lambda$CDM and $H_0^{\mathrm{R22}}$ value. Different colours represent different data sets.}
    \label{fig:Radar_H0_R22}
  \end{minipage}
  \caption{Distances measured in standard deviations ($\sigma$ units) between the constrained $H_0$ values obtained from the $f_{1-3}$CDM models together with $\Lambda$CDM and the $H_0^{\mathrm{P18}}$ value on the left-hand side and $H_0^{\mathrm{R22}}$ on the right-hand side. Different colours represent different data sets.}
    \label{fig:Radar_H0_P18_R22}
\end{figure} 
Table~\ref{tab:CB_CBf_AIC_BIC} highlights that the CC+BAO dataset generally exhibits a lower $\chi^2_\mathrm{min}$, implying a better fit without RSD. However, closer inspection of the $\Delta$AIC values for CC+BAO+RSD reveals a significant reduction, indicating the increased favorability of this dataset in certain cases. It's noteworthy that the BIC values for CC+BAO+RSD are slightly higher due to more data points. Regarding model performance, the $f_2$ model consistently presents the lowest $\Delta$AIC and $\Delta$BIC values, signifying its reliable performance. The $f_1$ and $f_3$ models also perform well, showing suitability for this dataset. In contrast, the $f_4$CDM model tends to lean towards the $\Lambda$CDM model, as suggested by higher $\Delta$AIC and $\Delta$BIC values, indicating weaker data support for $f_4$CDM.

Table~\ref{tab:Pf_CPB_CPBf_AIC_BIC} reveals intriguing findings. Starting with the $\mathrm{PN}^+$\,\&\,SH0ES+RSD dataset, the models are favored over $\Lambda$CDM, as reflected in negative AIC and BIC values, indicating a preference for the models. While this preference is not consistent across all models for the CC+$\mathrm{PN}^+$\,\&\,SH0ES+BAO dataset, it reemerges for CC+$\mathrm{PN}^+$\,\&\,SH0ES+BAO+RSD, especially pronounced for the $f_2$ and $f_3$ models. This suggests that, for these two models within the CC+$\mathrm{PN}^+$\,\&\,SH0ES+BAO+RSD datasets, the data leans towards favoring these models over $\Lambda$CDM. However, it is important to note that the evidence is not strong enough to definitively favor these models over $\Lambda$CDM. The BIC values, although not negative, are nearly zero, indicating an inconclusive preference for either model. Conversely, the $f_4$ model shows strong disfavor and preference for $\Lambda$CDM, prompting its exclusion from further analysis, shifting our focus to the first three models.

The previous analysis is further supported by Fig.~\ref{fig:Radar_H0_LCDM}. In this figure, we compare the values of $H_0$ obtained from $f_i(T)$ ($i={1,2,3}$) models to the values of $H_0$ obtained from the $\Lambda$CDM model for the same data sets, as shown in Appendix~\ref{sec:LCDM}. This visualization illustrates the variations in $H_0$ across different data sets, expressed in terms of $\sigma$ units, with each dataset represented by a distinct color. For each $f(T)$ model, we observe that the $H_0$ values fall within $1 \sigma$ of the corresponding $\Lambda$CDM values. Therefore, the $H_0$ values obtained for the different data sets are consistent with those of $\Lambda$CDM.

The discrepancy between the locally measured expansion rate of the Universe and the values inferred from observations of the CMB has prompted us to extend our previous analysis. We aim to investigate how these models perform with different values of $H_0$, specifically considering the P18 value $H^{\mathrm{P18}}_0 = 67.4 \pm 0.5 {\rm\, km \,s}^{-1} {\rm Mpc}^{-1}$ \cite{Planck:2018vyg} and R22 value of $H^{\mathrm{R22}}_0 = 73.30 \pm 1.4 {\rm\, km \, s}^{-1} {\rm Mpc}^{-1}$ \cite{Riess:2021jrx}, as shown in Fig.~\ref{fig:Radar_H0_P18_R22} (in both cases we include the $\Lambda$CDM values from Appendix~\ref{sec:LCDM}). 

In Fig.~\ref{fig:Radar_H0_P18_R22}, on the left-hand-side, we observe that the $H_0$ values are within 3.5$\sigma$ of the P18 value, except for the $\mathrm{PN}^+$\,\&\,SH0ES data set. However, in the figure, we also include values from RSD data only. This high $\sigma$ value for the $\mathrm{PN}^+$\,\&\,SH0ES data is expected, as it dominates and is included within the R22 value itself. Similarly, in Fig.~\ref{fig:Radar_H0_P18_R22}, on the right-hand side, we also see that the $H_0$ values are within approximately 3.5$\sigma$ of the R22 value. This suggests that the $H_0$ values obtained from our analysis fall approximately midway between the two quoted values of P18 and R22.

\begin{figure}
    \centering
    \includegraphics[width = \textwidth]{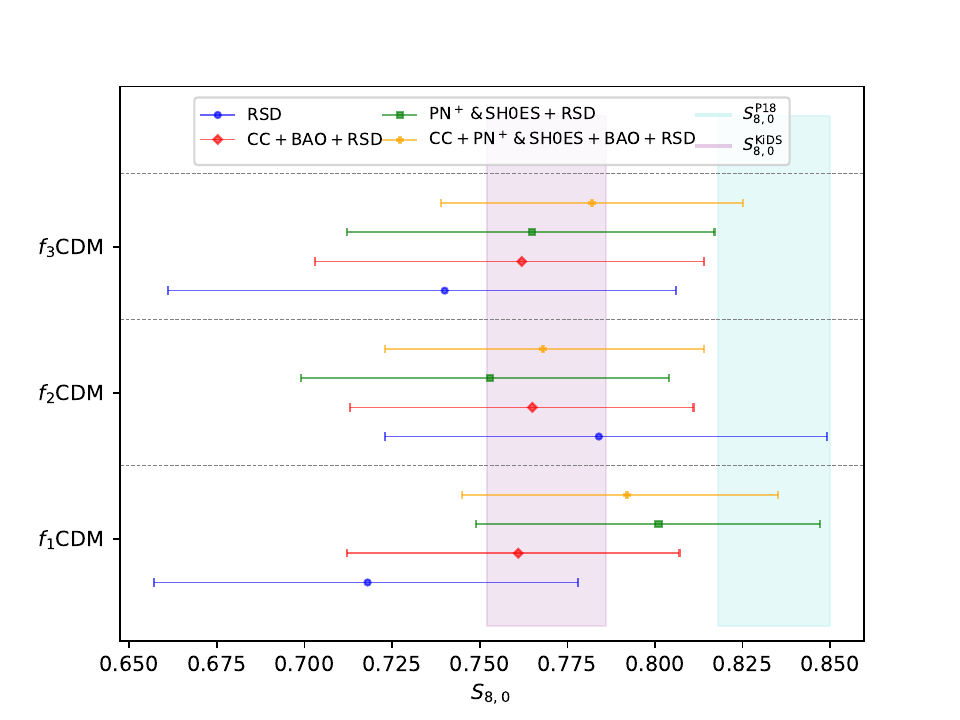}
    \caption{A whisker plot illustrating the constrained values of $S_{8,0}$ as derived from the $f_{1-3}$CDM models. The cyan and purple vertical bands depict the respective $1\sigma$ ranges of $S_{8,0}^{\mathrm{P18}}$ and $S_{8,0}^{\mathrm{KiDS}}$, whilst the colored error bars illustrate the inferred model dependent $1\sigma$ constraints from each respective data set.}
    \label{fig:S80_whisker_plot}
\end{figure}

\begin{figure}
    \centering
    \includegraphics[width = 0.7\textwidth]{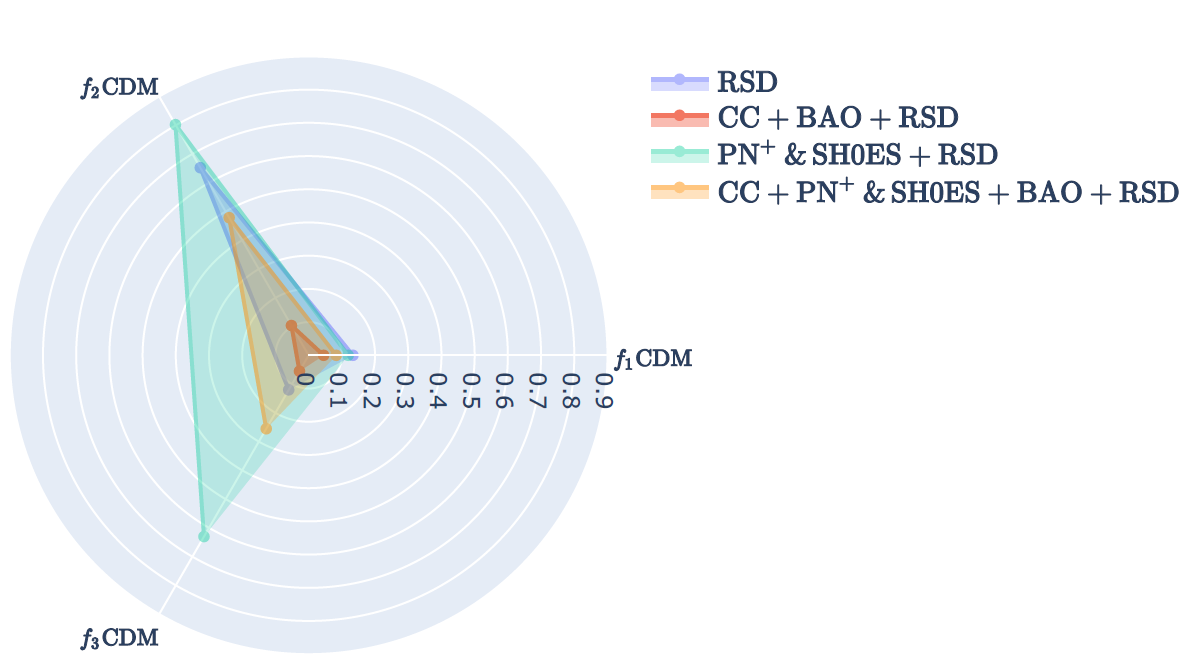}
    \caption{Distances measured in standard deviations ($\sigma$ units) between the constrained $S_{8,0}$ values obtained from the $f_{1-3}$CDM models and their corresponding values in the $\Lambda$CDM model. Different colours represent different data sets.}
    \label{fig:Radar_S80_LCDM}
\end{figure}

\begin{figure}[t]
    \hspace{-1.2cm}
  \begin{minipage}{0.56\textwidth}
    \includegraphics[width = \textwidth]{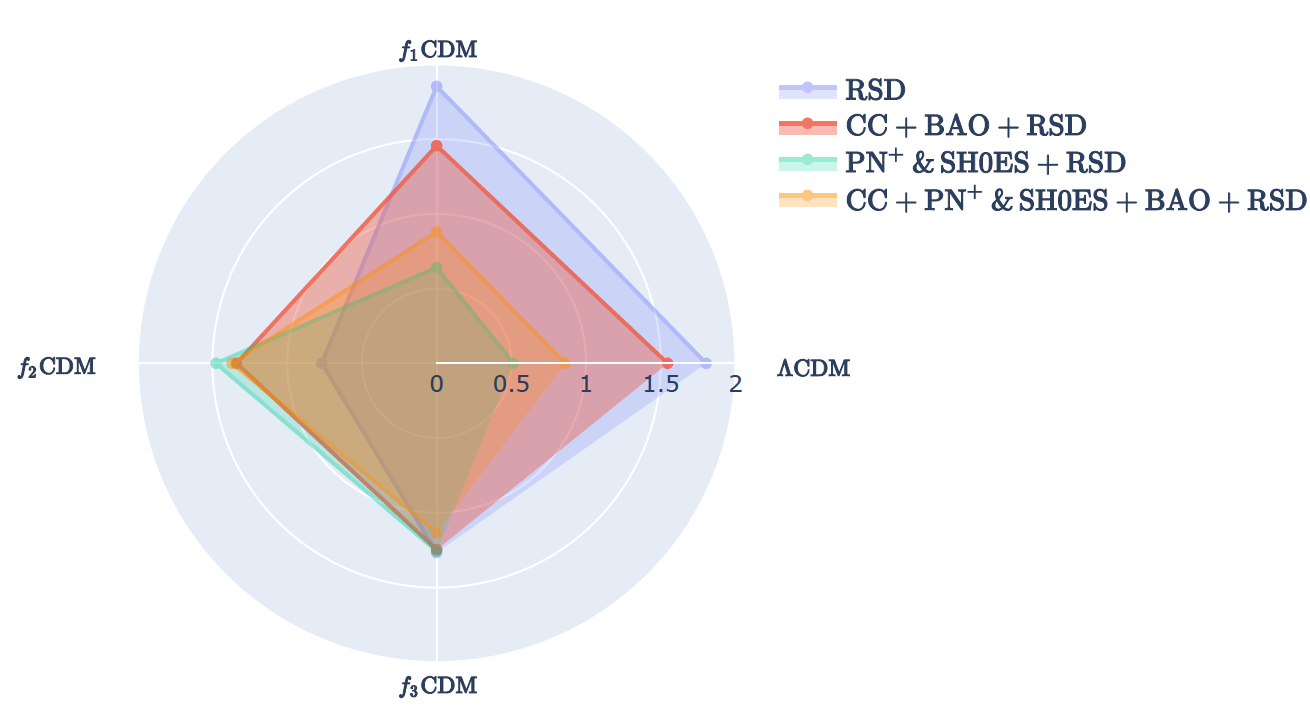}
    % \captionof{figure}{Distances measured in standard deviations ($\sigma$ units) between the constrained $S_{8,0}$ values obtained from the $f_{1-3}$CDM models together with $\Lambda$CDM and $S_{8,0}^{\mathrm{P18}}$ value. Different colours represent different data sets.}
    \label{fig:Radar_S80_P18}
  \end{minipage}
  \begin{minipage}{0.56\textwidth}
 \includegraphics[width = \textwidth]{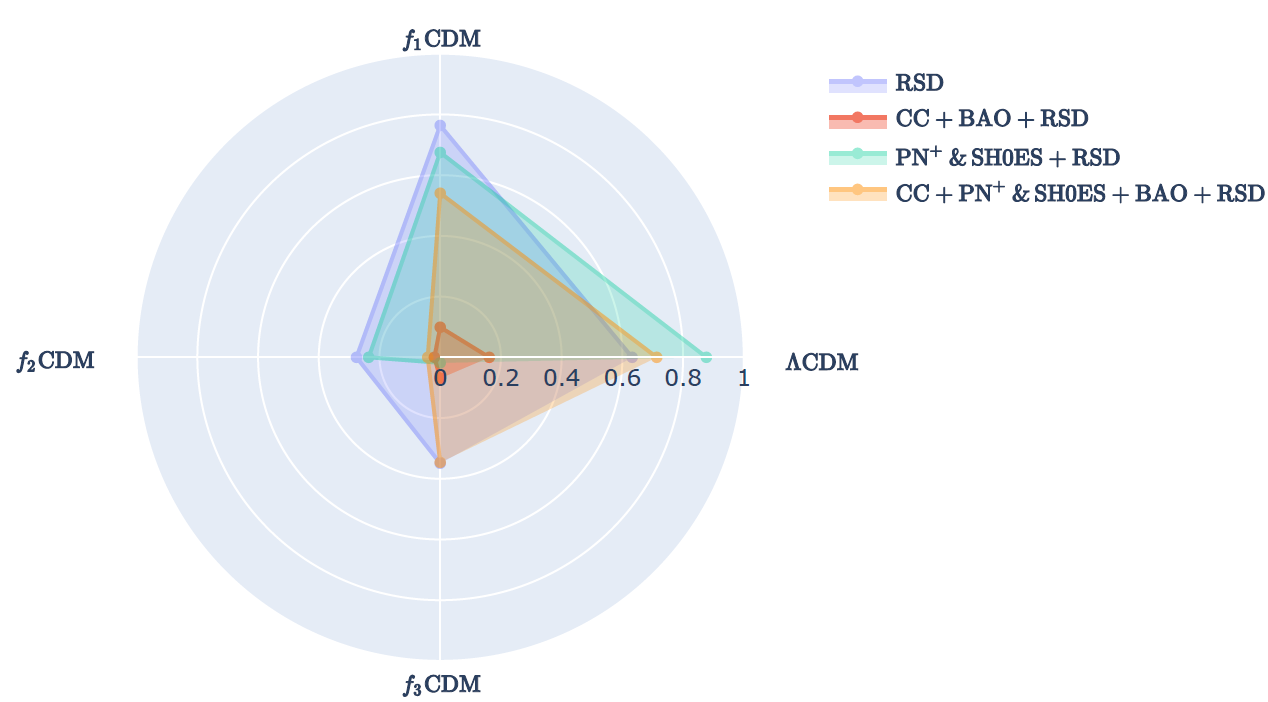}
    % \captionof{figure}{Distances measured in standard deviations ($\sigma$ units) between the constrained $S_{8,0}$ values obtained from the $f_{1-3}$CDM models together with $\Lambda$CDM and $S_{8,0}^{\mathrm{KiDS}}$ value. Different colours represent different data sets.}
    \label{fig:Radar_S80_KiDS}
  \end{minipage}
  \caption{Distances measured in standard deviations ($\sigma$ units) between the constrained $S_{8,0}$ values obtained from the $f_{1-3}$CDM models together with $\Lambda$CDM and the $S_{8,0}^{\mathrm{P18}}$ value on the left-hand side and $S_{8,0}^{\mathrm{KiDS}}$ on the right-hand side. Different colours represent different data sets.}
  \label{fig:Radar_S80_P18_KiDS}
\end{figure}

Additionally, we present a comprehensive comparison of our $S_{8,0}$ values with the Planck Collaboration's $S_{8,0}^{\mathrm{P18}} = 0.834 \pm 0.016$ \cite{Planck:2018vyg} and KiDS-1000 value $S_{8,0}^{\mathrm{KiDS}} = 0.766^{+0.020}{-0.014}$ \cite{KiDS:2020suj}. The results are summarized in Fig.~\ref{fig:S80_whisker_plot}, highlighting the impact of different datasets on our models.  a comparison between $S{8,0}$ values in our models and those in $\Lambda$CDM is presented in Fig.\ref{fig:Radar_S80_LCDM}. Furthermore, Fig.\ref{fig:Radar_S80_P18_KiDS} provides a comparison of $S_{8,0}^{\mathrm{P18}}$ and $S_{8,0}^{\mathrm{KiDS}}$ with values derived from $f_{1-3}$CDM models in $\sigma$ units. When compared to $S_{8,0}^{\mathrm{P18}}$, on the left-hand side of same figure, we find a maximum discrepancy of $2\sigma$ between the values derived from $\Lambda$CDM and the $f(T)$ models, and this is primarily observed when considering the RSD data set in isolation. However, as more data sets are included, particularly CC+$\mathrm{PN}^+$\,\&\,SH0ES+BAO+RSD, which is the most comprehensive data set in our analysis, this discrepancy diminishes.

On the right pane of Fig.~\ref{fig:Radar_S80_P18_KiDS}, where we compare with $S_{8,0}^{\mathrm{KiDS}}$, the discrepancy is further reduced to $1\sigma$, indicating that the values obtained from our analysis are closer to the KiDS-1000 measurements than the Planck values. Notably, CC+$\mathrm{PN}^+$\,\&\,SH0ES+BAO+RSD performs well in this regard, significantly reducing the $\sigma$ values. Consequently, it appears that with the combination of the available data sets and the $f_{1-3}(T)$ models, the tensions between the CMB and locally determined values are reduced, especially for $S_{8,0}$.

\section{Conclusion}\label{sec:conclusion}

The most popular models in the literature of TG have been probed in this work against the latest expansion data as well as RSD measurements which has expanded the constraint profile of each of these models, as well as their impact on physically observable cosmological parameters. For the local data being used, we had a combination of CC, $\mathrm{PN}^+$\,\&\,SH0ES and BAO data. Besides being the most populated data sets and highest precision, these data sets give a range of points across redshift space. In particular, the $\mathrm{PN}^+$\,\&\,SH0ES sample is the largest SNIa in the literature and the principal late-time indicator of the cosmic tensions problem in the Hubble constant. We use RSD data since it is sensitive to the growth of structure formation and so can be used to constrain the perturbative sector of the models under consideration. In each of these models and for each data set combination, we performed a full MCMC analysis obtaining constraints on all the cosmological parameters. Additionally, we compared the performance of each model for each data set combination against the standard $\Lambda$CDM model using the $\chi^2_\mathrm{min}$, AIC, and BIC statistical indicators. Given the increasing tension being reported by various studies on the value of the Hubble constant, we study its constrained value, but also consider the effect of a changing Hubble constant will have on the parameters related to the growth of structure formation, namely $\sigma_{8,0}$ and $S_{8,0}$.

For reference purposes, we provide the $\Lambda$CDM constraint values for each of the data set combinations we consider. This is done in Appendix~\ref{sec:LCDM}, which is important for estimating the statistical indicators in the model sections. Generally, our analysis shows consistency with $\Lambda$CDM but some differences do arise, which may show further distinction as the precision in measurements improves. As one might expect, the strongest data set in the analysis was the $\mathrm{PN}^+$\,\&\,SH0ES sample since it consistently drastically reduces the statistical error for any baseline data set. On the other hand, the statistical metrics show that expansion data CC+$\mathrm{PN}^+$\,\&\,SH0ES+BAO combination provides evidence for $\Lambda$CDM while when the RSD data set is included, there is marginal preference for the $f(T)$ model under consideration. Moreover, they offer best fits on cosmic parameters that are more aligned with a higher Hubble constant. As for the $S_{8,0}$ parameter, the models are largely consistent with the most recent reported values in the literature. 

This analysis offers precision insights into the behavior of these models, but also on TG more generally, when using combinations of this expansion profile and RSD data sets. The study suggests that these models may offer some promise as toy models for modified cosmological scenario model building. We intend to use these observations to extend this analysis to include CMB power spectra as well as other early Universe data sets to more fully assess the competitiveness of these models.

\section*{Acknowledgements}\label{sec:acknowledgements}
The authors would like to acknowledge support from the Malta Digital Innovation Authority through the IntelliVerse grant. 
CE-R acknowledges the Royal Astronomical Society as FRAS 10147 and is supported by PAPIIT UNAM Project TA100122. The authors would also like to acknowledge funding from ``The Malta Council for Science and Technology'' as part of the ``FUSION R\&I: Research Excellence Programme'' REP-2023-019 (CosmoLearn) Project.
This research has been carried out using computational facilities procured through the European Regional Development Fund, Project No. ERDF-080 ``A supercomputing laboratory for the University of Malta''. This paper is based upon work from COST Action CA21136 {\it Addressing observational tensions in cosmology with systematics and fundamental physics} (CosmoVerse) supported by COST (European Cooperation in Science and Technology). JLS would also like to acknowledge funding from “The Malta Council for Science and Technology” in project IPAS-2023-010.

%%%%%%%%%%%%%%%%%%%%%%%%%%%%%%%%%%%%%%%%%

\appendix
\section{Model parameter \texorpdfstring{$p_i$}{pi} versus \texorpdfstring{$S_{8,0}$}{S80} plots} \label{sec:App_pvsS}

In this section, we present the posteriors together with their confidence regions of the $p_i$ and $S_{8,0}$ parameters to investigate the correlation between the two. 

\begin{figure}[]
    \centering
    \includegraphics[width = 0.45\textwidth]{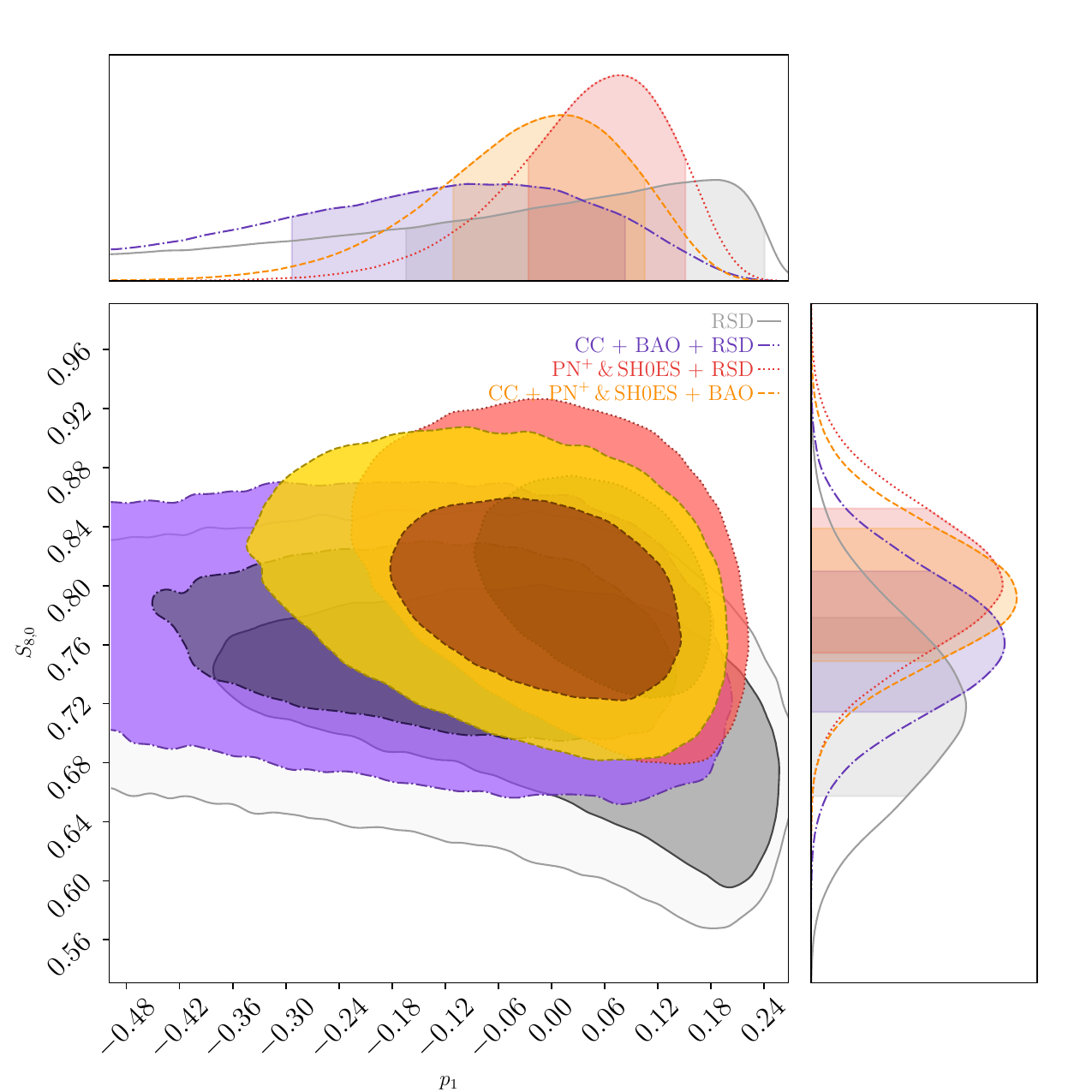}
     \includegraphics[width = 0.45\textwidth]{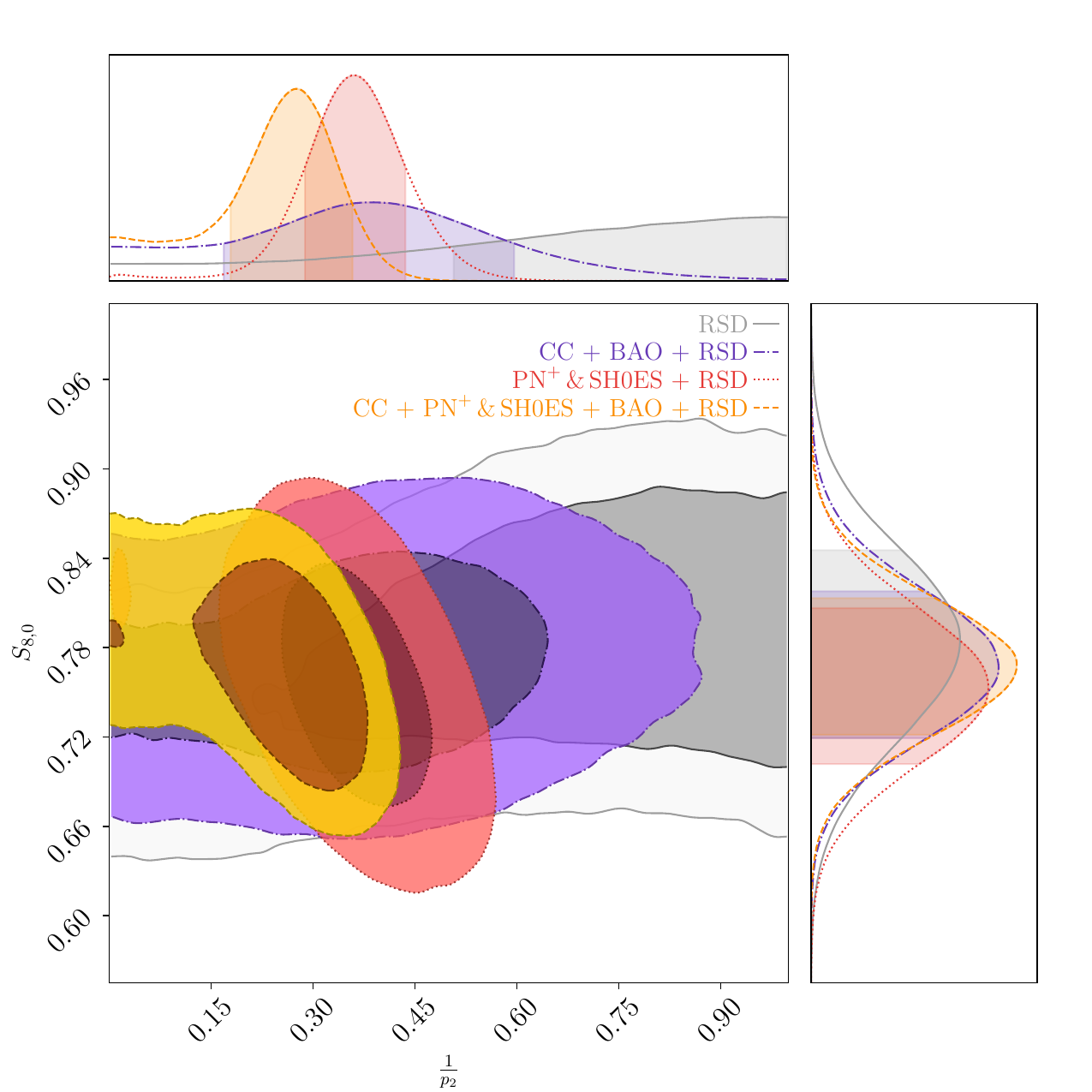}
     \includegraphics[width = 0.45\textwidth]{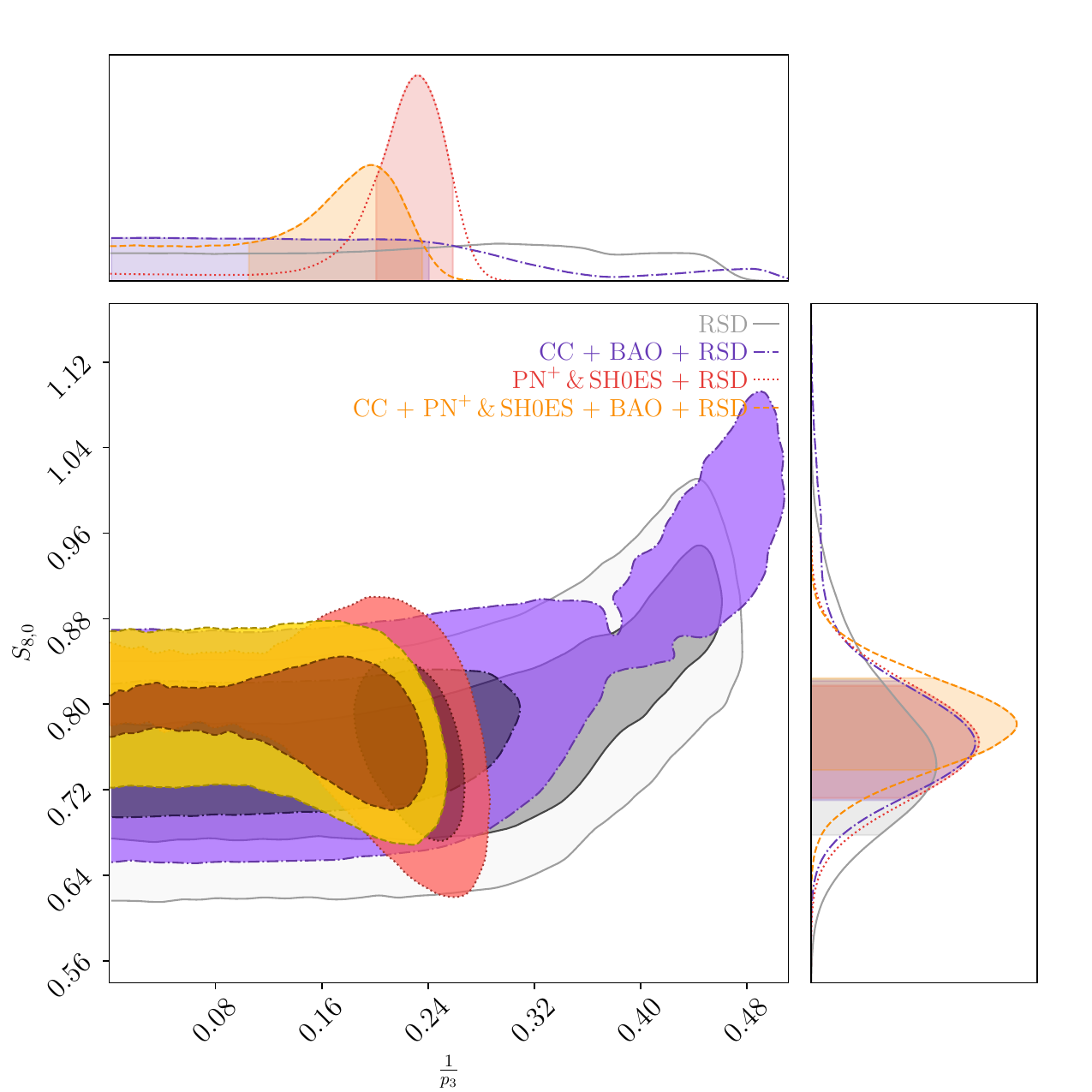}
     \includegraphics[width = 0.45\textwidth]{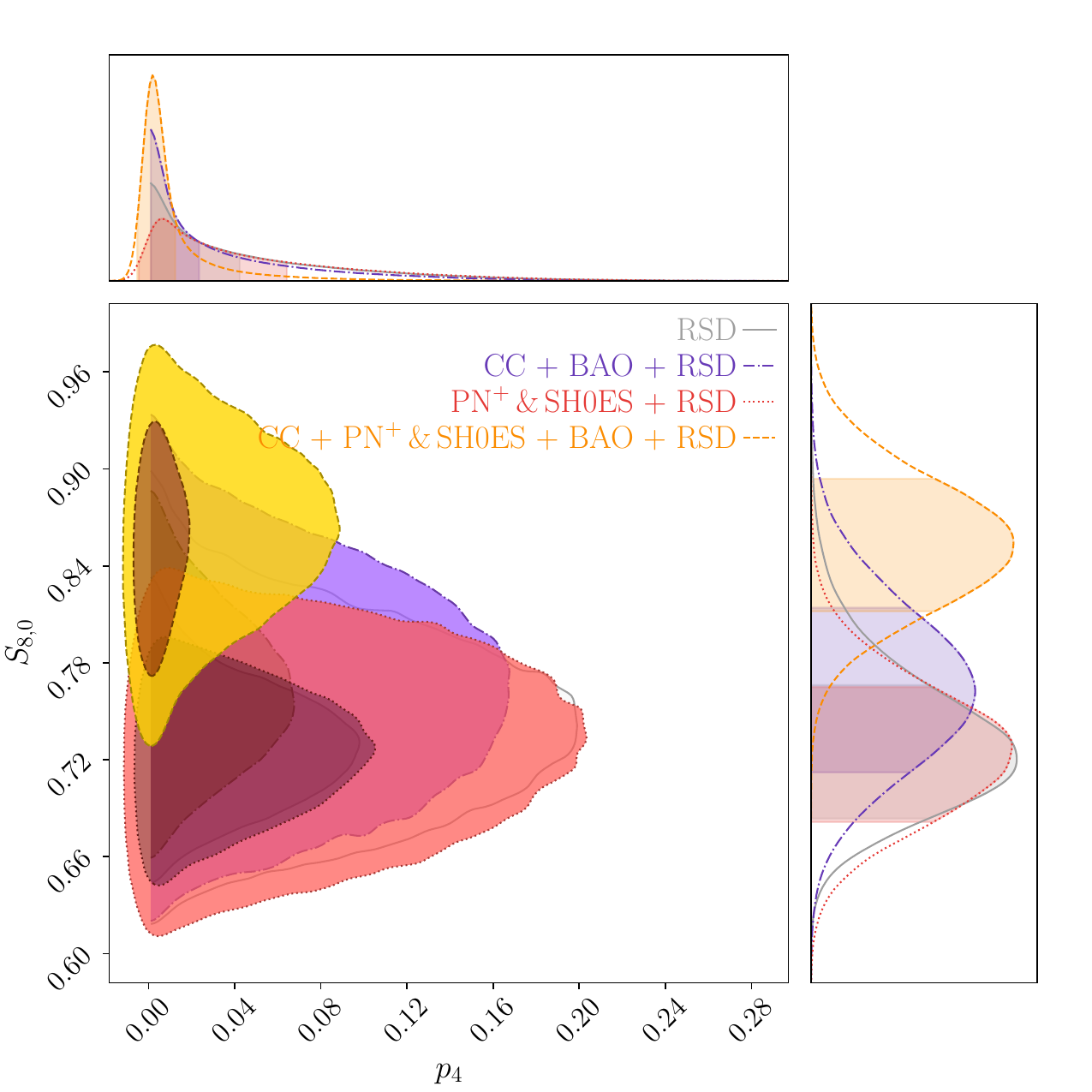}
    \caption{\textit{Top left:} Contour plots showing the relationship between the $p_1$ parameter and the $S_{8,0}$ parameter for the $f_1$CDM model (Power Law Model).
   \textit{Top right:} Contour plots showing the relationship between the $\frac{1}{p_2}$ parameter and the $S_{8,0}$ parameter for the $f_2$CDM model (Linder Model).
   \textit{Left bottom:} Contour plots showing the relationship between the $\frac{1}{p_3}$ parameter and the $S_{8,0}$ parameter for the $f_1$CDM model (Exponential Model).
   \textit{Right bottom:} Contour plots showing the relationship between the $p_4$ parameter and the $S_{8,0}$ parameter for the $f_1$CDM model (Logarithmic Model).
    }
    \label{fig:PLM_p_vs_S80}
\end{figure}

%\begin{figure}[h]
 %   \centering
  %  \includegraphics[width = 0.5\textwidth]{Figures/LM_p_vs_S80_final.pdf}
   % \caption{Contour plots showing the relationship between the $\frac{1}{p_2}$ parameter and the $S_{8,0}$ parameter for the $f_2$CDM model (Linder Model).}
    %\label{fig:LM_p_vs_S80}
%\end{figure}

%\begin{figure}[h]
%    \centering
 %   \includegraphics[width = 0.5\textwidth]{Figures/LM2_p_vs_S80_final.pdf}
  %  \caption{Contour plots showing the relationship between the $\frac{1}{p_3}$ parameter and the $S_{8,0}$ parameter for the $f_1$CDM model (Exponential Model).}
  %  \label{fig:LM2_p_vs_S80}
%\end{figure}

%\begin{figure}[h]
%    \centering
 %   \includegraphics[width = 0.5\textwidth]{Figures/LogM_p_vs_S80_final.pdf}
 %   \caption{Contour plots showing the relationship between the $p_4$ parameter and the $S_{8,0}$ parameter for the $f_1$CDM model (Logarithmic Model).}
 %   \label{fig:LogM_p_vs_S80}
%\end{figure}

%%%%%%%%%%%%%%%%%%%%%%%%%%%%%%%%%%%%
\clearpage
\section{\texorpdfstring{$\Lambda$}{Lambda}CDM Model}\label{sec:LCDM}

\begin{figure}[]
    \centering
    \includegraphics[width = 0.8\textwidth]{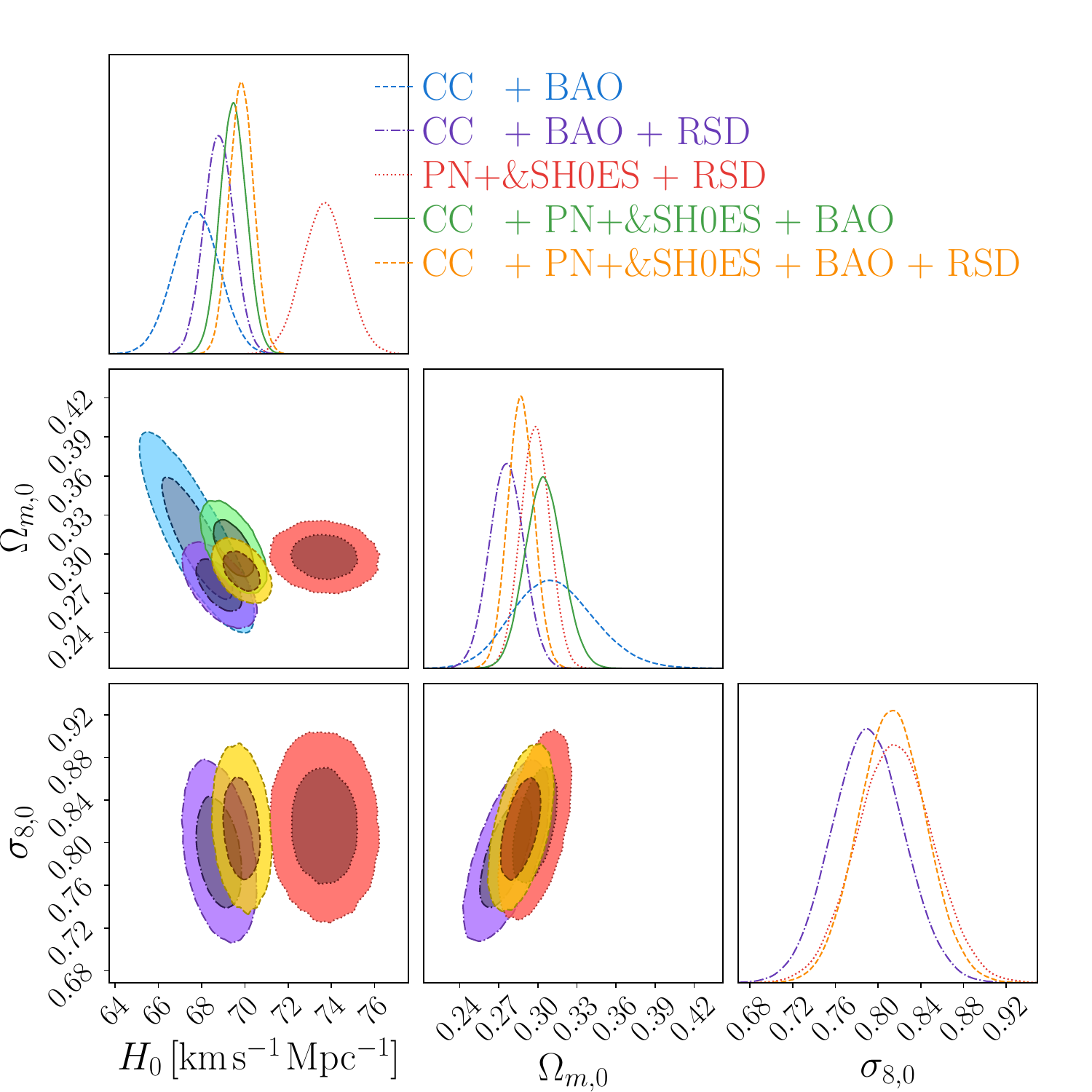}
    \caption{Confidence contours and posterior distributions for the $\Lambda$CDM model (Logarithmic Model) parameters, including $H_0$ and $\Omega_{m,0}$. In cases where the RSD data is incorporated (purple, red, and yellow contours), the $\sigma_{8,0}$ parameter is also displayed.}
    \label{fig:LCDM}
\end{figure}

In Section~\ref{sec:analysis}, we provide comparisons between the models and their corresponding $\Lambda$CDM values. Here, we present the posterior distributions and confidence regions of the $\Lambda$CDM model in Figure~\ref{fig:LCDM} and provide additional details and precise values in Table~\ref{tab:LCDM}.

\begin{table*}
\resizebox{\textwidth}{!}{%
\centering  
 %   \hspace{-2cm}
    \begin{tabular}{ccccc}
        \hline
		Data Sets & $H_0 \mathrm{\hspace{0.15cm}[km \hspace{0.1cm} s^{-1} \hspace{0.1cm}Mpc ^{-1}]}$ & $\Omega_{m,0}$ & $\sigma_{8,0}$ & $M$ \\ 
		\hline
		CC \, + BAO  & $67.8\pm 1.1$ & $0.308^{+0.032}_{-0.029}$ & -- & --  \\ 
		CC \, +  BAO + RSD & $68.77^{+0.71}_{-0.67}$ & $0.276^{+0.013}_{-0.014}$ & $0.789^{+0.035}_{-0.033}$ & --  \\ 
		$\mathrm{PN}^+$\,\&\,SH0ES + RSD & $73.71^{+0.97}_{-1.06}$ & $0.298^{+0.011}_{-0.012}$ & $0.814^{+0.037}_{-0.034}$ & $-19.249^{+0.027}_{-0.032}$ \\ 
		CC \, + $\mathrm{PN}^+$\,\&\,SH0ES + BAO  & $69.47^{+0.59}_{-0.63}$ & $0.304^{+0.015}_{-0.014}$ & -- &  $-19.375\pm 0.017$ \\ 
		CC \, + $\mathrm{PN}^+$\,\&\,SH0ES + BAO + RSD & $69.84^{+0.55}_{-0.56}$ & $0.288^{+0.010}_{-0.010}$ & $0.815^{+0.030}_{-0.033}$ & $-19.369^{+0.016}_{-0.017}$  \\ 
		\hline
    \end{tabular}
    }
      \caption{Exact results for $\Lambda$CDM model that include the parameters $H_0$ and $\Omega_{m,0}$. The $\sigma_{8,0}$ parameter and the nuisance parameter $M$, are provided for data sets that include RSD or $\mathrm{PN}^+$\,\&\,SH0ES, respectively  otherwise, they are left empty.}
    \label{tab:LCDM}
\end{table*}

\begin{figure}[t]
  \begin{minipage}{0.5\textwidth}
    \captionsetup{width=0.9\textwidth}
    \includegraphics[width = \textwidth]{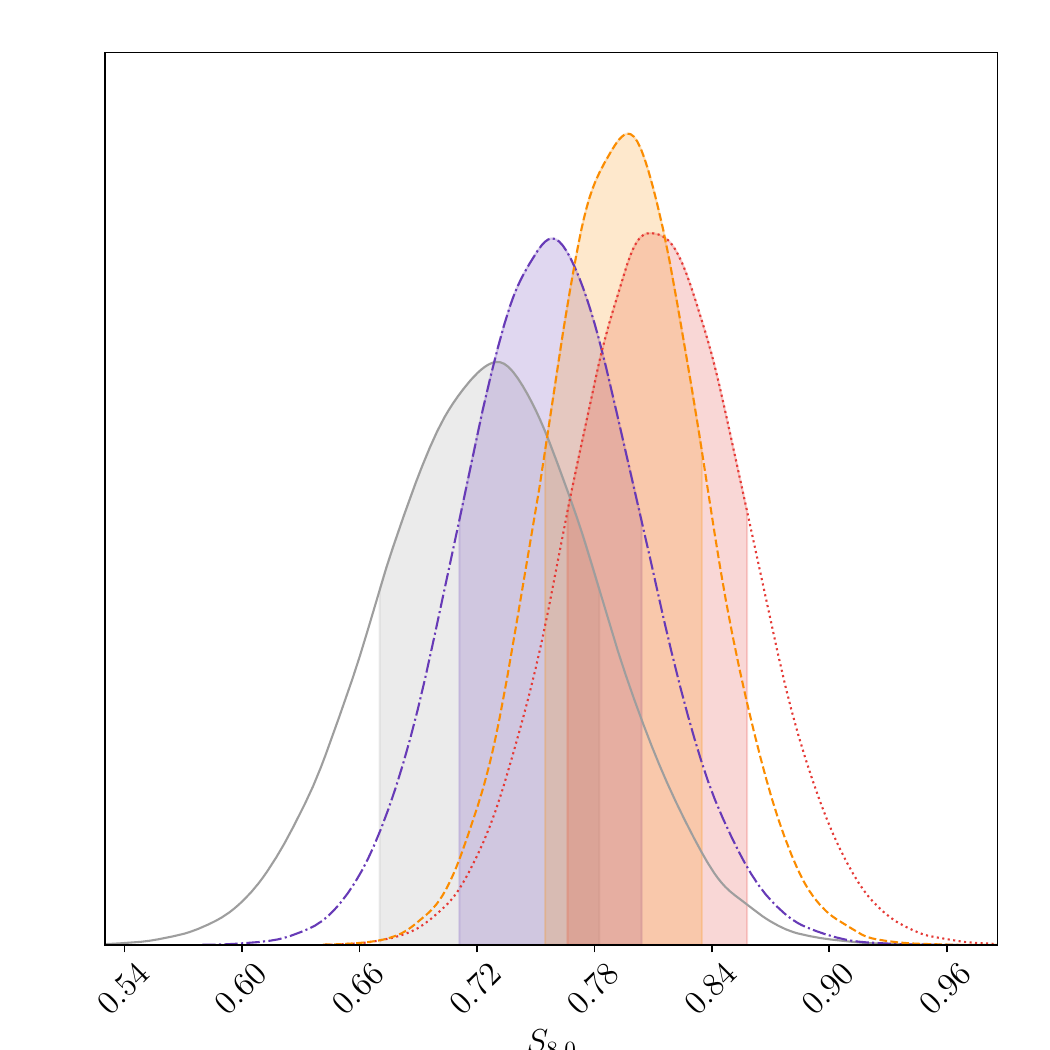}
    \captionof{figure}{Posterior distribution for the $S_{8,0}$ parameter in the $\Lambda$CDM model. Legend: Grey denotes the RSD data, purple corresponds to CC+BAO+RSD data, red represents the $\mathrm{PN}^+$\,\&,SH0ES + RSD dataset, while orange indicates CC + $\mathrm{PN}^+$\,\&\,SH0ES + BAO + RSD data.}
    \label{fig:LCDM_S80}
  \end{minipage}
  \begin{minipage}{0.5\textwidth}
    \captionsetup{width=0.9\textwidth}
    \begin{tabular}{lc}
        \hline
		Model & $S_{8,0}$ \\ 
		\hline
		 \textcolor{mygray}{\tikz[baseline=-0.75ex]\draw [thick,solid] (0,0) -- (0.5,0); RSD} & $0.729^{+0.053}_{-0.059}$ \\ 
		\texttransparent{0.7}{\textcolor{patriarch}{\tikz[baseline=-0.75ex]\draw [thick,dash dot dot] (0,0) -- (0.5,0);CC +  BAO + RSD}} & $0.758^{+0.046}_{-0.047}$ \\ 
		\texttransparent{0.7}{\textcolor{myred}{\tikz[baseline=-0.75ex]\draw [thick,dotted] (0,0) -- (0.5,0);$\mathrm{PN}^+$\,\&\,SH0ES + RSD}} & $0.809^{+0.050}_{-0.042}$ \\ 
		\texttransparent{0.8}{\textcolor{myorange}{\tikz[baseline=-0.75ex]\draw [thick,dashed] (0,0) -- (0.5,0);CC + $\mathrm{PN}^+$\,\&\,SH0ES + BAO + RSD}} & $0.797^{+0.038}_{-0.042}$ \\ 
		\hline
    \end{tabular}
    \captionof{table}{Exact $S_{8,0}$ values corresponding to various data sets for the $\Lambda$CDM model.}
    \label{tab:LCDM_S80}
  \end{minipage}
\end{figure}

% \begin{figure}[h]
%     \centering
%     \includegraphics[width = 0.5\textwidth]{Figures/LCDM_Om0_vs_S80.pdf}
%     \caption{LCDM}
%     \label{fig:LCDM_Om0_vs_S80}
% \end{figure}

\clearpage
\bibliographystyle{JHEP}
\bibliography{references}

\end{document}